\documentclass[fleqn,useAMS,usenatbib]{mnras}
\usepackage{soul}
\usepackage{ragged2e}
\usepackage[usenames]{xcolor}

\definecolor{green2}{rgb}{0.2,0.7,0.1}
\definecolor{rust}{rgb}{0.7,0.1,0.1}

\usepackage{adjustbox}
\usepackage{tablefootnote}
\usepackage{orcidlink}
\usepackage{longtable}
\usepackage[super]{nth}
\usepackage{hyperref}
\usepackage{stfloats} 
\usepackage[caption=false]{subfig} 
\DeclareRobustCommand{\VAN}[3]{#2} 
\title[SN~2000ch]
{Repeating periodic eruptions of the supernova impostor SN~2000ch}

\author[M. Aghakhanloo et al.]{Mojgan Aghakhanloo \orcidlink{0000-0001-8341-3940},$^{1}$\thanks{E-mail:
aghakhanloo@arizona.edu} Nathan Smith \orcidlink{0000-0001-5510-2424},$^1$ 
Peter Milne,$^1$ Jennifer E. Andrews \orcidlink{0000-0003-0123-0062},$^{2}$ \newauthor
Alexei V. Filippenko \orcidlink{0000-0003-3460-0103},$^{3}$ 
Jacob E. Jencson \orcidlink{0000-0001-5754-4007},$^{4}$ 
David J. Sand \orcidlink{0000-0003-4102-380X},$^{1}$ 
Schuyler D. Van Dyk \orcidlink{0000-0001-9038-9950},$^{5}$ \newauthor
Samuel Wyatt \orcidlink{0000-0003-2732-4956},$^{6}$ 
WeiKang Zheng \orcidlink{0000-0002-2636-6508}$^{3}$\\ 
$^1$ Steward Observatory, University of Arizona, 933 N. Cherry Ave., Tucson, AZ 85721, USA  \\ 
$^2$ Gemini Observatory, 670 N. Aohoku Place, Hilo, Hawaii, 96720, USA \\  
$^3$ Department of Astronomy, University of California, Berkeley, CA 94720-3411, USA\\
$^4$ The Johns Hopkins University, Baltimore, MD 21218, USA\\
$^5$ Caltech/IPAC, Mailcode 100-22, Pasadena, CA 91125, USA\\
$^6$ Department of Physics, University of Washington, Seattle, WA, USA}

\begin{document}
\pagerange{\pageref{firstpage}--\pageref{lastpage}} \pubyear{2023}
\maketitle
\label{firstpage}

\begin{abstract}
We analyse photometric observations of the supernova (SN) impostor SN~2000ch in NGC~3432 covering the time since its discovery. This source was previously observed to have four outbursts in 2000--2010. Observations now reveal at least three additional outbursts in 2004--2007, and sixteen outbursts in 2010--2022. Outburst light curves are irregular and multipeaked, exhibiting a wide variety of peak magnitude, duration, and shape. The outbursts after 2008 repeat with a period of $200.7\pm{2}$~d, while the outburst in 2000 seems to match with a shorter period. The next outburst should occur around January/February 2023.  We propose that these periodic eruptions arise from violent interaction around times of periastron in an eccentric binary system, similar to the periastron encounters of $\eta$ Carinae leading up to its Great Eruption, and resembling the erratic pre-SN eruptions of SN~2009ip. We attribute the irregularity of the eruptions to the interplay between the orbit and the variability of the luminous blue variable (LBV) primary star, wherein each successive periastron pass may have a different intensity or duration due to the changing radius and mass-loss rate of the LBV-like primary.   Such outbursts may occasionally be weak or undetectable if the LBV is relatively quiescent at periastron, but can be much more extreme when the LBV is active. The observed change in orbital period may be a consequence of mass lost in outbursts. Given the similarity to the progenitor of SN~2009ip, SN~2000ch deserves continued attention in the event it is headed for a stellar merger or a SN-like explosion.

\end{abstract}


\section{INTRODUCTION}\label{sec:intro}
\begin{figure}
\includegraphics[width=0.47\textwidth]{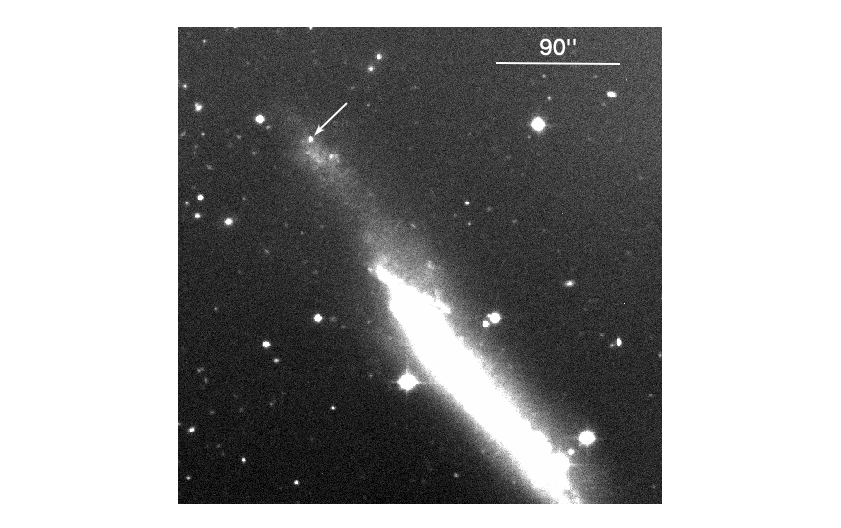}
\caption{Kuiper $R$-band image of SN~2000ch obtained on 2018 March 20. The position of SN~2000ch is marked with an arrow. North is up and east is to the left.}\label{fig:ds9image}
\end{figure}

The supernova (SN) impostor SN~2000ch (also known as NGC 3432-LBV1 and AT~2000ch) was first discovered on 2000 May 3 (UT dates are used throughout this paper) by \cite{P00} during the course of the Lick Observatory Supernova Search \citep{L02,F01}. SN~2000ch is located about 123{\arcsec} east and 180{\arcsec} north of the centre of the spiral galaxy NGC~3432 (see Fig.~\ref{fig:ds9image}). SN~2000ch has experienced a series of outbursts, starting with its original brightening in 2000 \citep{W04}, followed by three more outbursts observed in 2008 and 2009 \citep{P10}. These outbursts of SN~2000ch tend to show rapid brightening and fading, sometimes flickering repeatedly over short timescales of days to weeks during a single eruptive episode.  The outbursts do not look like typical decade-long S~Doradus cycles \citep{vg01,sterken08} of luminous blue variables (LBVs), but instead more closely resemble the flickering of SN~1954J before its giant eruption \citep{S10,S11}, the peaks of $\eta$~Carinae leading up to its 19th-century Great Eruption \citep{sf11}, or the brief repeating peaks of SN~2009ip before its SN explosion \citep{S10,P13}.  The repetition of rapid brightening and fading suggests violent binary encounters as a possible underlying mechanism \citep{P10,SN11,S11}. 

Visual-wavelength spectra of SN~2000ch also differ from typical spectra of LBVs.  When in eruption, LBVs usually show relatively cool temperatures, with a mix of absorption, emission, and P~Cygni lines due to a moderately cool, low-ionisation, dense wind, with line widths that indicate relatively slow outflow speeds of 100--500 km s$^{-1}$.  By contrast, spectra of SN~2000ch exhibit a very smooth blue continuum with strong and broad emission from He~{\sc i} and H~{\sc i} Balmer lines, where these emission lines have Lorentzian-shaped profiles with widths of 1500--2000 km s$^{-1}$ \citep{P10,S11}. Spectra of SN~2000ch more closely resemble those of H-rich Wolf-Rayet (WR) stars rather than normal LBVs.  \citet{S10} and \citet{B18} noted that the pre-SN outbursts of SN~2009ip and SN 2015bh had spectra that were very similar to those of SN~2000ch. Further spectral analysis of SN~2000ch is beyond the scope of this paper; in a subsequent paper, we will present spectroscopic observations of SN~2000ch.

SN~2000ch is classified as part of a group of eruptive transients known as 
``SN impostors," which are usually associated with evolved massive stars \citep{V00,S11}. SN impostors are characterised by their strong photometric variability and presumably nonterminal eruptions during their lifetime. Their eruptions are often interpreted as giant eruptions of LBVs, but they show diverse properties in their outburst spectra, fading rates, peak magnitudes, and other observable properties \citep{S11}. The most famous Galactic SN impostors are P~Cygni and $\eta$~Car, which are prototypes for giant LBV eruptions \citep{H47,D69,D88}.   Many mechanisms have been proposed to explain these LBV-like eruptions, including super-Eddington winds \citep{O04,S06}, envelope instability \citep{G93,G99}, and binary interaction of various sorts \citep{S04,PP10,SN11,S14}.  \citet{S11} provides a list of SN impostors and discusses candidate mechanisms that may explain phenomena associated with them.

Another famous example of an SN impostor is SN~2009ip in NGC 7259, which suffered several LBV-like outbursts before its SN explosion \citep{S10,P13}. 
SN2009ip's LBV-like eruptions culminated in a very bright Type IIn SN-like event\footnote{\citep[See][for a review of SN spectral classification]{F97}.} in 2012 \citep{M13}, providing strong evidence for LBVs as possible progenitors of SNe~IIn. 
Although some authors proposed that the 2012 eruption of SN~2009ip was a nonterminal SN event \citep[e.g.,][]{K13,P13,fraser15} recent evidence confirms that this was indeed a terminal SN explosion \citep{S22}. Similar to recent observations of SN~2009ip, late-time observations of SN 2015bh \citep{J22} and AT 2016jbu \citep{B22} also suggest terminal explosions, where the LBV-like progenitors are now gone. Later, we will discuss similar rebrightening peaks observed in SN~2000ch's photometric evolution.  SN~2000ch's similarity to SN~2009ip's pre-SN eruptions hints that SN~2000ch might also be headed for a catastrophe, making it a target of great interest.  

In the specific case of SN~2000ch, the cause of its eruptions and the properties of the stellar system have remained an unsolved problem. \citet{W04} noted that the erratic and rapid variability of SN~2000ch was very unusual among known transients and variables, but suggested connections to massive LBVs and SN impostors like $\eta$~Car and SN~1997bs \citep{V00}.  SN~2000ch's outbursts are brighter than a typical LBV outburst, but not as bright as $\eta$~Car’s giant eruption. \cite{P10} suggested that SN~2000ch's outbursts may be due to repeated mass ejection episodes of LBVs, or perhaps interaction with a massive, binary companion. They noted some similarities between SN~2000ch and the WR/LBV multiple system HD 5980 in the Small Magellanic Cloud. A merger in a triple system that leaves behind a binary LBV+WR system is a scenario that was proposed to explain the giant eruption of $\eta$~Car \citep{S18}. If variations observed in SN~2000ch's light curve are caused by interaction in a binary system, (quasi)-periodic structures should exist in the light curve. However, \cite{P10} concluded that the modulation of SN~2000ch's light curve in data available at that time was not regular enough to claim that the SN~2000ch's outbursts are periodic.

In this paper, we present the continued photometric evolution of SN~2000ch using various facilities. The main goal of this paper is to document major outbursts of SN~2000ch after the December 2009 event and search for any degree of periodicity. First, in Section~\ref{sec:obs}, we describe observations and the data-reduction process. In Section~\ref{sec:LC}, we analyse the previous and recent major outbursts of SN~2000ch. Section~\ref{sec:Priodicity} discusses the photometric evolution of SN~2000ch to investigate its possible periodicity. Furthermore, using the observed period, we predict the times of future events.

\section{NEW OBSERVATIONS}\label{sec:obs}
\subsection{New Photometry}
We obtained optical photometry of SN~2000ch across many epochs using the 0.76~m Katzman Automatic Imaging Telescope \citep[KAIT;][]{F01} at Lick Observatory, the 0.6~m robotic  Super-LOTIS telescope \citep[Livermore Optical Transient Imaging System;][]{W08} on Kitt Peak, and the Kuiper 61~inch telescope on Mt. Bigelow, Arizona. To supplement these, we also retrieved photometry from the Zwicky Transient Facility \citep[ZTF;][]{B19, G19, M19} and ATLAS \citep[Asteroid Terrestrial-impact Last Alert System;][]{T18}, as described below. 

Images of SN~2000ch were obtained from 1999-02-24 to 2021-12-11 by KAIT.  These included data from the original discovery and the first outburst of SN~2000ch reported by \citet{W04}.  We remeasured the KAIT photometry using the original images. Remeasurements agree with the original data within about 0.1 mag. The unfiltered KAIT photometry of SN~2000ch is summarised in Table~\ref{tab:KAIT}. Unfiltered KAIT magnitudes are similar to the standard $R$ passband \citep{L03}. They are calibrated to local stars, with their magnitudes transformed into the Landolt $R$-band system\footnote{\url{https://classic.sdss.org/dr7/algorithms/sdssUBVRITransform.html\#Lupton2005}}. SN~2000ch is in a relatively isolated outer part of its host galaxy with low background and
no strong gradient in background light level. Point-spread-function (PSF) photometry was obtained using DAOPHOT \citep{S87} from the IDL Astronomy User’s Library\footnote{\url{http://idlastro.gsfc.nasa.gov/}}, and was calibrated using photometry of field stars in the same image as SN~2000ch. Note that if the detection magnitude is fainter than the limiting magnitude (lim~mag) in Table~\ref{tab:KAIT}, the point is shown as an upper limit in light-curve plots below.

\begin{figure}
\includegraphics[width=0.45\textwidth]{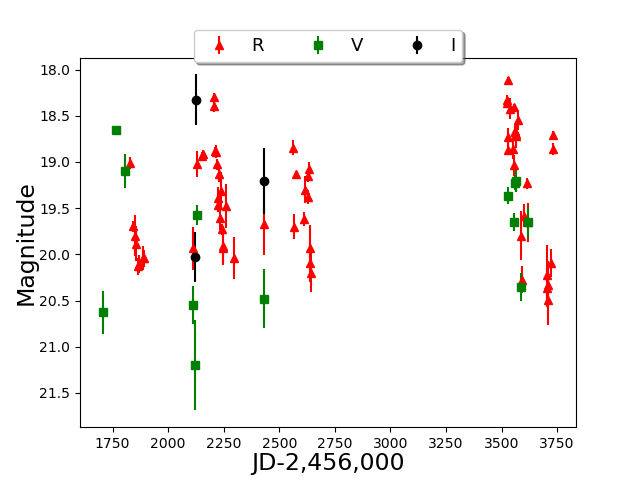}
\caption{Multiband light curve of SN~2000ch using Super-LOTIS photometry. Data are listed in Table~\ref{tab:SuperLOTIS}.}\label{fig:SuperLOTIS}
\end{figure}

Super-LOTIS observations were obtained from 2016 November 12 up to the present epoch. Standard image reduction was performed using a semi-automatic routine. Photometric measurements were extracted using standard multi-aperture photometry techniques and were calibrated using photometry of field stars in the same image as SN~2000ch.  Super-LOTIS optical magnitudes and associated uncertainties are given in Table~\ref{tab:SuperLOTIS}. Fig.~\ref{fig:SuperLOTIS} shows the multiband light curve of SN~2000ch observed by Super-LOTIS. In addition, 32 epochs of SN~2000ch were acquired between 2013-01-19 and 2022-05-08,  using the MONT4K CCD imager on the Kuiper telescope. Kuiper data were reduced with the 
same methods and standard-star magnitudes as employed for the Super-LOTIS data. We present the Kuiper photometry of SN~2000ch in Table~\ref{tab:Kuiper}.

We also obtained photometry from publicly available archives.  The position of SN~2000ch was repeatedly observed by the ZTF public surveys. ZTF public data are available online\footnote{\raggedright\url{https://irsa.ipac.caltech.edu/Missions/ztf.html}}.  We collect the sources within the 2\arcsec position of SN~2000ch from ZTF DR15. As recommended by the ZTF Science Data System Explanatory Supplement (ZSDS)\footnote{\raggedright\url{http://web.ipac.caltech.edu/staff/fmasci/ztf/ztf_pipelines_deliverables.pdf}}, we use data with {\it catflags}=0 to exclude the poor-quality images. The {\it catflags} filter reduced the number of data points from 391 to 366. 
SN~2000ch was also observed repeatedly by the ATLAS Project, using the cyan and orange filters ($c$-ATLAS and $o$-ATLAS, respectively).  ATLAS data are available in the ATLAS Forced Photometry server\footnote{\raggedright\url{https://fallingstar-data.com/forcedphot/}}. Since $o$-ATLAS has better temporal coverage and is closer to the red filters from other telescopes noted above, we only use $o$-ATLAS data in our analysis. Employing a Python script by David Young\footnote{\raggedright\url{https://gist.github.com/thespacedoctor/86777fa5a9567b7939e8d84fd8cf6a76}}, sigma-clipping of the ATLAS data was implemented to remove rogue data points.  In addition, we exclude the data points that have negative fluxes, and when magnitude uncertainties are larger than 0.5 mag. In the next subsection, we use the new photometric data along with previous data in the literature to analyse the continued photometric evolution of SN~2000ch.

\section{The Light Curve}\label{sec:LC}

The full light curve of SN~2000ch from all sources is shown in Fig.~\ref{fig:AllLC}. Fig.~\ref{fig:AllLCA} displays the photometric evolution of SN~2000ch from 1994 to 2010 including measurements from the literature. Fig.~\ref{fig:AllLCB} continues in time, showing the light curve of SN~2000ch from 2010 to 2022 using our new photometry.   Observations in Sloan filters (from \citealt{P10} and \citealt{Muller}) are scaled to Johnson-Bessell $R$ by applying a shift of $-0.18$. Similarly, ZTF observations are also shifted by $-$0.18, assuming that the SDSS $r$ filter is the closest match to the ZTF $r$ filter. The $o$-ATLAS filter roughly corresponds to SDSS $r+i$, but because the ATLAS filters are nonstandard, we do not attempt to transform them to other magnitude systems. 
Since our goal is to look for periodicity in the relative brightness rather than deriving absolute physical properties, these shifts are merely a convenience for display in the light-curve plots.  

The solid vertical grey bands in Fig.~\ref{fig:AllLC} indicate the time interval from July 21 until October 1 each year, when SN~2000ch is difficult or impossible to observe owing to its position being too close to the Sun in the sky. Since the period we find below is less than a year, it is expected that some outbursts fall within these unobservable windows or at a boundary. SN~2000ch shows a baseline brightness around 19.5 mag, but has many episodes where it briefly brightens by 1--2 mag, and it occasionally dips to 20--21 mag, as in the brief eclipse-like event following the bright peak in 2000 originally reported by \citet{W04}.  In the following, we first review the previous four major outbursts of SN~2000ch that are documented in the literature, and then we explore the detailed light curve of SN~2000ch since 2010 using our new photometry to document the more recent outbursts.

\subsection{Previously Documented Outbursts}\label{sec:PreviousLC}
Fig.~\ref{fig:WagnerA} shows the detailed evolution of SN~2000ch during its first recorded  outburst. On 2000 May 3, SN~2000ch reached maximum brightness, with $R=17.5$ mag.  Photometry reported by \cite{W04} start from 2000 April 10, and the first left-most point in Fig.~\ref{fig:WagnerA} represents the pre-outburst brightness. After the main brightness peak, there are additional fainter peaks in the light curve and a sharp drop in brightness, which those authors interpreted as a result of post-outburst dust formation or an eclipse phase. We also notice minor inconsistencies (marked in grey) between Fig.~\ref{fig:WagnerA} and the figure presented by \cite{W04}. These inconsistencies are probably due to typos in the reported data. 

\begin{figure*}
\subfloat[]{
  \includegraphics[width=0.85\textwidth]{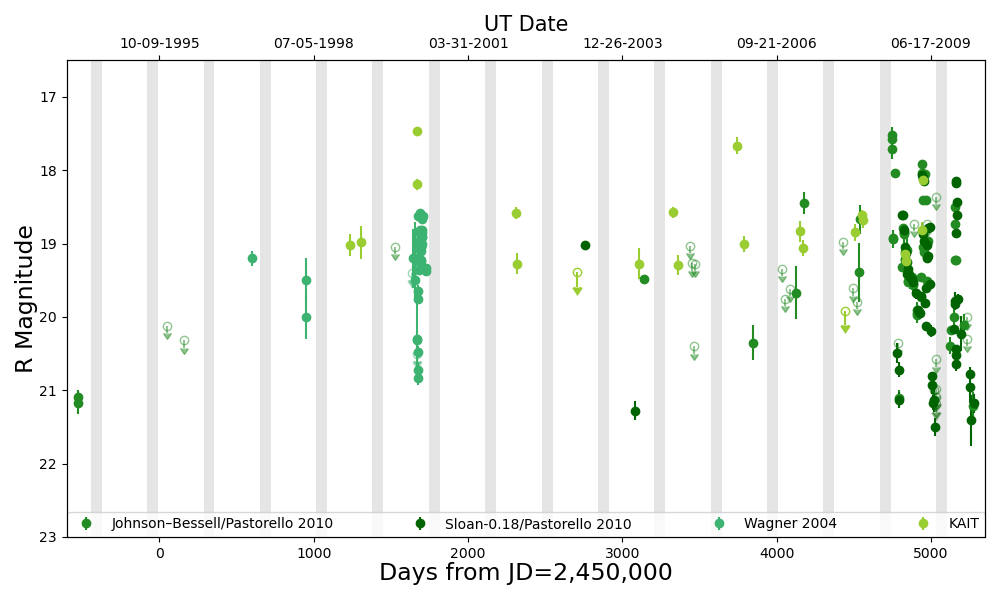}
  \label{fig:AllLCA}}\\
\subfloat[]{
  \includegraphics[width=0.85\textwidth]{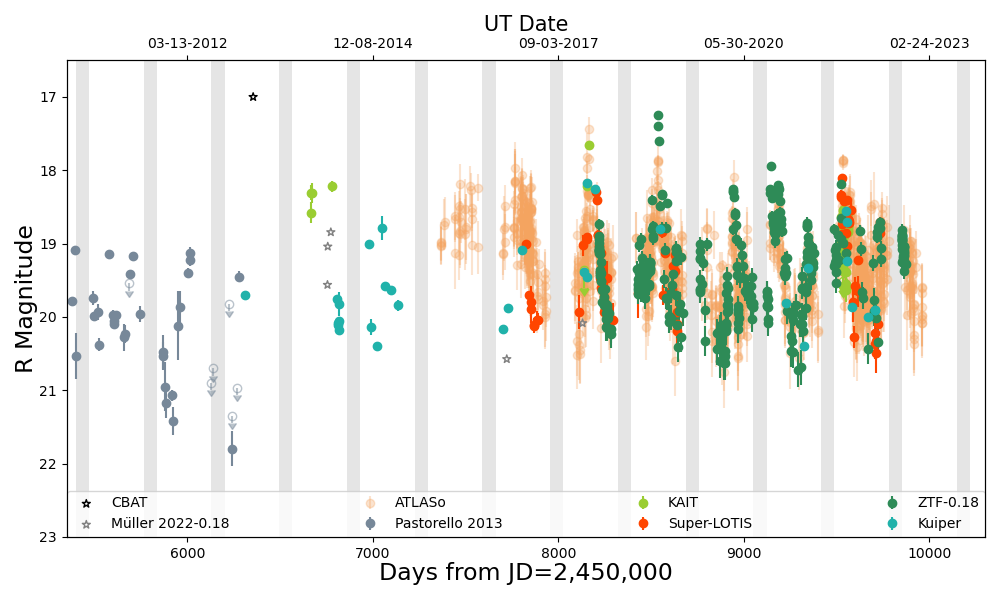}
  \label{fig:AllLCB}}
  
\caption{The observed light curve of SN~2000ch, which is mostly in the $R$-band filter, but ATLAS data are in the ``orange'' ($o$) filter and CBAT data are unfiltered. The top panel shows the photometric evolution of SN~2000ch from 1994 to 2010. The 2000 data are from \protect\cite{W04}, while the 2008 and 2009 data are from \protect\cite{P10}. The bottom panel shows the light curve from 2010 to 2022 using our new photometry. In both panels, grey areas indicate times each year from July 21 until October 1, when SN~2000ch was difficult to observe owing to its proximity to the Sun. Photometric observations cover at least twenty-three outbursts over the past two decades. See Figs.~\ref{fig:Wagner}, \ref{fig:PastorelloZoomed}, \ref{fig:LC2010to14}, and \ref{fig:RecentZoomed} that show zoomed-in  portions of the light curve.}\label{fig:AllLC}
\end{figure*}

\begin{figure}
\subfloat[]{
  \includegraphics[width=0.45\textwidth]{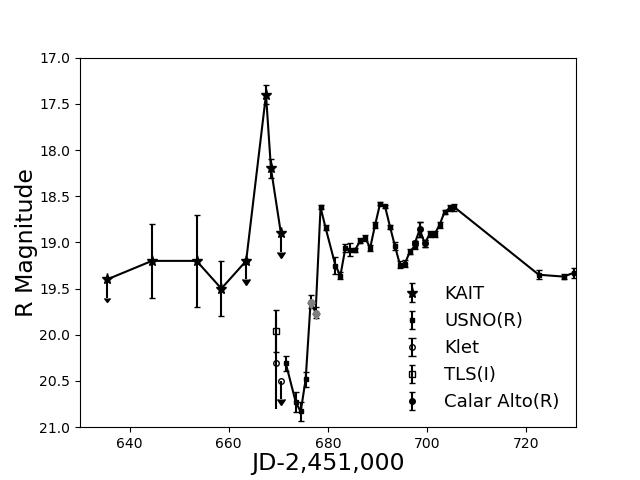}
  \label{fig:WagnerA}
}
\newline
\subfloat[]{
  \includegraphics[width=0.45\textwidth]{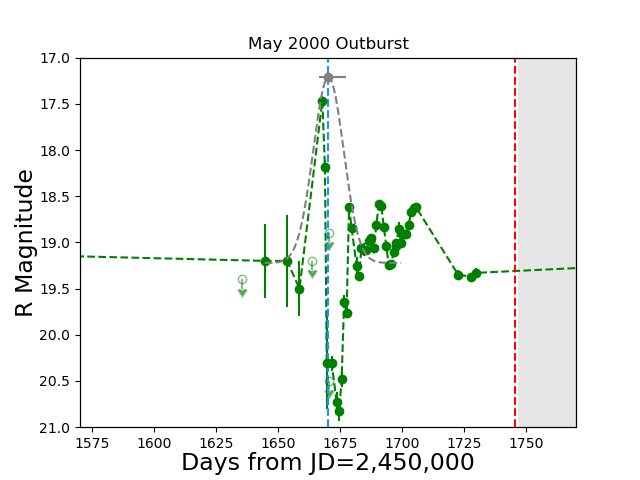}
  \label{fig:WagnerB}
}
\caption{The top panel (a) shows the light curve of SN~2000ch during its first outburst on 2000 May 3. We reproduce the light curve using the data from \protect\cite{W04}. The grey circles show the minor inconsistencies with the light curve published  by \protect\cite{W04}. The bottom panel (b) shows the $R$-band light curve by including our remeasured KAIT photometry.  The dashed grey  curve and the circle indicate the overall Gaussian comparison and the adopted centre of activity, respectively. The bottom panel shows a larger time axis compared to the top panel for better comparison with other outbursts in Figs.~\ref{fig:PastorelloZoomed} and \ref{fig:RecentZoomed}. As in Fig.~\ref{fig:AllLC} and several figures that follow, the grey shaded area shows when SN~2000ch was difficult to observe. The red and blue vertical dashed lines mark the reference epoch, showing where the centre of the activity is predicted to be using the detected periods of 200.7 and 194 days, respectively, derived from many repeated eruptions (see Section~\ref{sec:Priodicity} for more details).} \label{fig:Wagner}
\end{figure}

\begin{figure}
\subfloat[]{
  \includegraphics[width=0.4\textwidth]{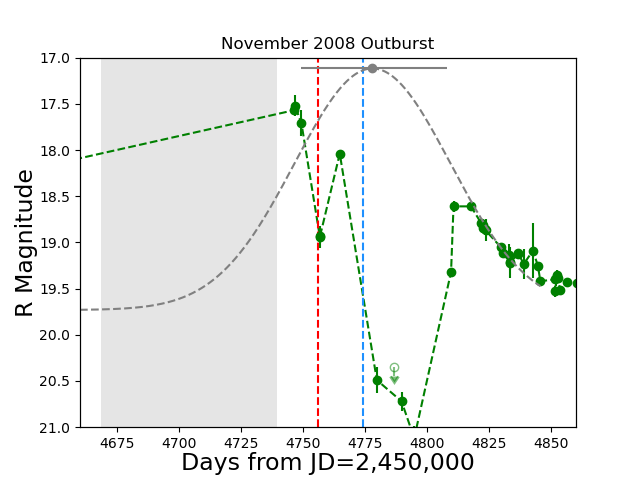}
  \label{fig:PastorelloZoomedA}
}
\newline
\subfloat[]{
  \includegraphics[width=0.4\textwidth]{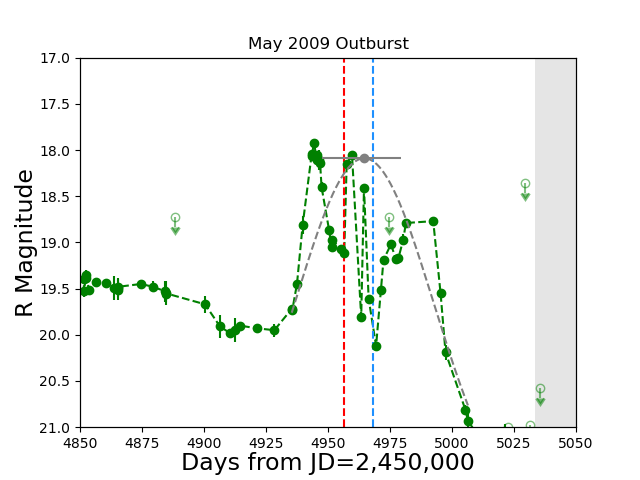}
  \label{fig:PastorelloZoomedB}
}
\newline
\subfloat[]{
  \includegraphics[width=0.4\textwidth]{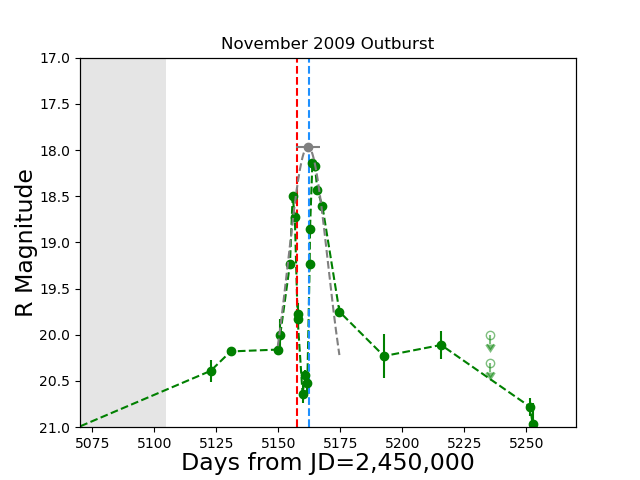}
  \label{fig:PastorelloZoomedC}
}
\caption{The $R$-band light curve of SN~2000ch during the 2008 and 2009 outbursts, including data from \citet{P10} and our KAIT photometry. The grey area indicates the interval when SN~2000ch was hard to observe, as in Fig.~\ref{fig:AllLC}. Dashed grey curves show Gaussian curves to each event and grey filled circles show centres of activity with error bars. The approximate times for the centre of activity are given at the top of the frames. If outbursts are partially blocked out by the star's dust or an eclipse, then we argue that the centre of activity is a better measure of the timing of the event than the time of peak (see text).}\label{fig:PastorelloZoomed}
\end{figure}

The second outburst of SN~2000ch on 2008 October 7 was documented by \cite{D08}. The 2008 outburst was first reported as a new transient, but later \cite{P10} found that the 2000 and 2008 events were both major outbursts associated with the same source, which was SN~2000ch. Fig.~\ref{fig:PastorelloZoomedA} shows the 2008 outburst.  In this event, data covering the rise and perhaps even the peak are missing because SN~2000ch was hard to observe during that time of year (grey shaded region). Based on the available data, SN~2000ch reached a peak of $R=17.5$ mag, and experienced oscillations in its brightness including a sharp drop to $\sim 21$ mag shortly after the main peak, similar to its first outburst in 2000.

The third and fourth outbursts of SN~2000ch in April and November 2009 (respectively) were reported by \cite{P10}. During the third and fourth outbursts, SN~2000ch reached peak $R$ magnitudes of 17.9 and 18.1, respectively (i.e., about 0.5--1.0 mag fainter than the first two events). Fig.~\ref{fig:PastorelloZoomedB} shows the third peak which is followed by multiple rapid peaks and dips. During the fourth outburst (Fig.~\ref{fig:PastorelloZoomedC}), SN~2000ch experienced two peaks within 8 d with approximately similar magnitudes. Those authors registered the second peak as the main outburst because it was slightly brighter. Later in Section~\ref{sec:Centroid}, we discuss that the outbursts have irregular variability and sometimes multiple closely spaced peaks of varying brightness, so the time of the brightest peak may not necessarily be the best indicator for the time of the outburst.

Fig.~\ref{fig:AllLCA} demonstrates that SN~2000ch experienced more outbursts after the 2000 event until 2008, which are not previously documented in the literature. We note dips in February 2002, November 2004, and February 2008. Unfortunately, data coverage is not good enough to study the detailed evolution of the light curve at these times. However, we should emphasise that in April 2004, SN~2000ch experienced a sharp brightening by $\sim 2$ mag. In January 2006, SN~2000ch also reached an apparent magnitude of 17.7, which is approximately as bright as the 2000 event, and then it faded by $\sim 3$ mag. Later, in March 2007, SN~2000ch again brightened by $\sim 1$ mag. In this work, we use the word ``outburst" for brightenings that are more than 1~mag. Therefore, SN~2000ch experienced at least three more outbursts before 2010, making a total of seven known outbursts in 2000--2010. 

\subsection{Post-2010 Outbursts}\label{sec:RecentLC}
Fig.~\ref{fig:AllLCB} shows the recent $R$-band light curve of SN~2000ch from 2010 to 2022. From 2010 through 2022, we find that SN~2000ch experienced at least sixteen more outbursts. All of SN~2000ch's outbursts are summarised in Table~\ref{tab:summaryoutbursts}. In the following, we describe each event in detail.  The designation for each event below corresponds to the date of the approximate midpoint of activity, which we discuss in the following section. We also note that since $o$-ATLAS data are in a different filter and many points are close to the limiting magnitude of that telescope ($\sim 19$ mag), they are shown in the background and they are not used to define a period (see Section~\ref{sec:Priodicity} for more details).

\begin{figure}
\includegraphics[width=0.45\textwidth]{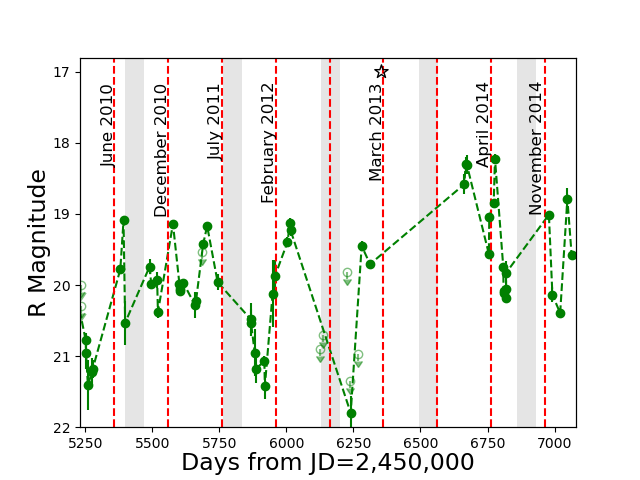}
\caption{The $R$-band light curve of SN~2000ch during the 2010-2014 outbursts. Data are from \citet{P13} and \citet{Muller}. Data also include our KAIT and Kuiper photometry. However, the data are sparse and the cadence is insufficient to match the overall Gaussian curves and estimate centres of activities.}\label{fig:LC2010to14}
\end{figure}

{\it 2010--2012:}  Additional photometry of SN~2000ch was published by \cite{P13} covering 2010--2013, although those authors did not discuss any additional outbursts seen in these data. Based on these data, we find that SN~2000ch experienced four additional outbursts in 2010--2012. Fig.~\ref{fig:LC2010to14} shows all four outbursts.

{\it March 2013:} In late 2012, SN~2000ch brightened by $\sim 2$ mag. Later, V. Nevski, E. Romas, and I. Molotov reported an outburst discovery of SN~2000ch using the ISON-Kislovodsk Observatory\footnote{\raggedright\url{http://www.cbat.eps.harvard.edu/unconf/followups/J10524126+3640086.html}}. In March 2013, SN~2000ch reached an apparent magnitude of $\sim 17.0$ (see Fig.~\ref{fig:AllLCB}). Given the period of 200.7 d (see Section~\ref{sec:Priodicity} for more details), the time of the reported discovery is within $\sim 7$ d of an expected event.

{\it April 2014:} Fig.~\ref{fig:LC2010to14} shows that in April 2014, SN~2000ch reached a peak at $R= 18.2$~mag. Since the submission of this paper, an independent work by \cite{Muller} also studied SN~2000ch. We include some of the photometry from that paper in our analysis (see Fig.~\ref{fig:AllLCB}). These data combined with our KAIT and Kuiper photometry show that SN~2000ch experienced another outburst in April 2014.

{\it November 2014:} Later in 2014, SN~2000ch faded by $\sim 1.6$ mag. The main peak is missing owing to poor data coverage before the dip. The data covering the seven outbursts from 2010 through 2014 are sparse and the cadence is insufficient to estimate the centroid of activity. It is unclear how well these broad outbursts line up with the predicted centre of activity (vertical dashed red line; see Section~\ref{sec:Centroid}) because of large gaps in the data.

\begin{figure*}
\begin{tabular}{ccc}
\subfloat[]{\label{fig:RecentZoomed1}
    \includegraphics[width=0.3\linewidth]{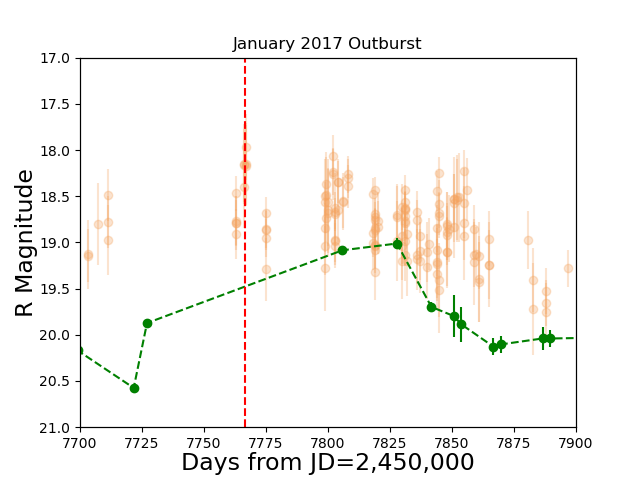}}&
    \subfloat[]{\label{fig:RecentZoomed2}
    \includegraphics[width=0.3\linewidth]{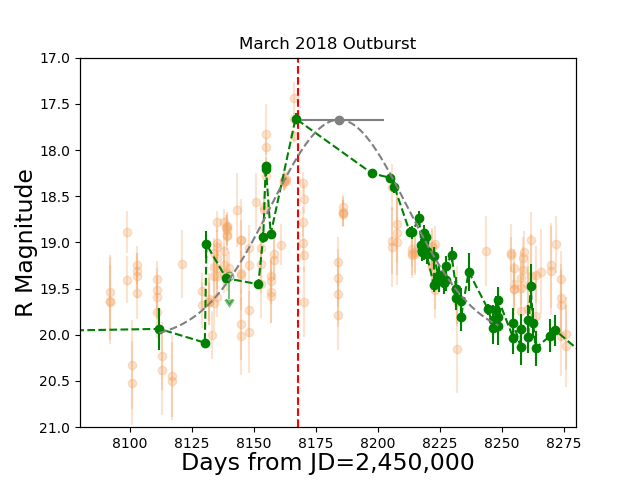}}&
    \subfloat[]{\label{fig:RecentZoomed3}
    \includegraphics[width=0.3\linewidth]{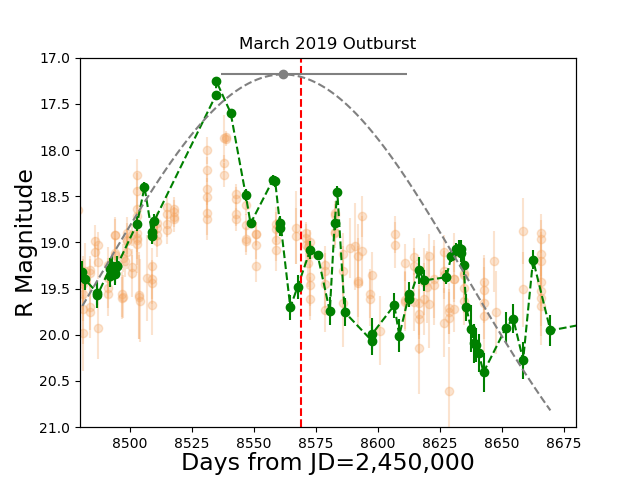}}\\
    \subfloat[]{\label{fig:RecentZoomed4}
    \includegraphics[width=0.3\linewidth]{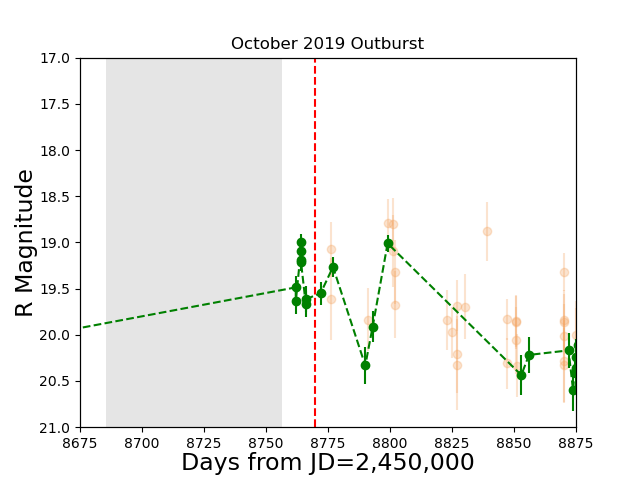}}&
    \subfloat[]{\label{fig:RecentZoomed5}
    \includegraphics[width=0.3\linewidth]{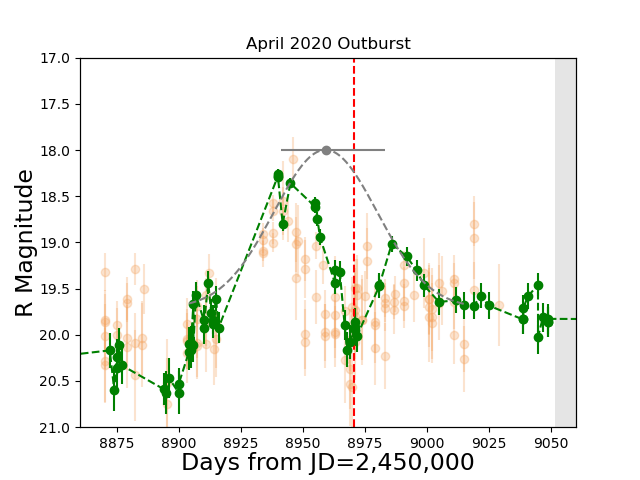}}&
    \subfloat[]{\label{fig:RecentZoomed6}
    \includegraphics[width=0.3\linewidth]{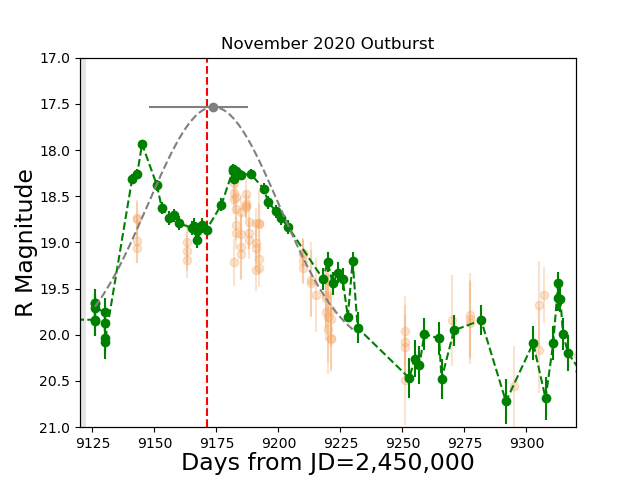}}\\
    \subfloat[]{\label{fig:RecentZoomed7}
        \includegraphics[width=0.3\linewidth]{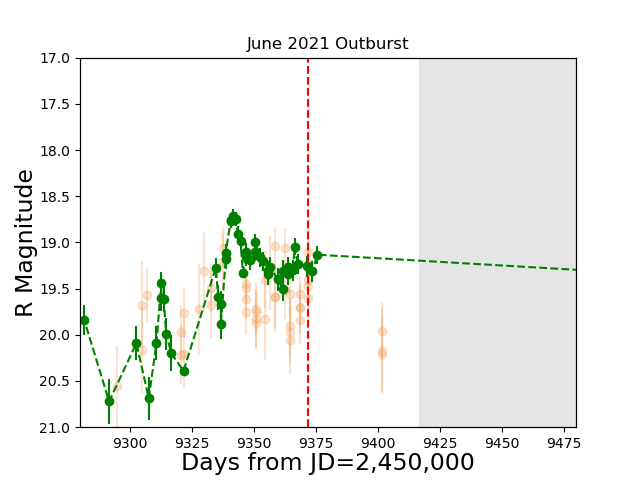}}&
        \subfloat[]{\label{fig:RecentZoomed8}
     \includegraphics[width=0.3\linewidth]{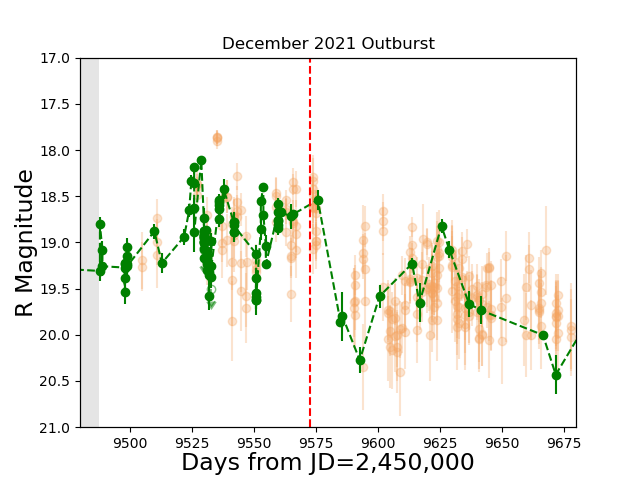}}&
     \subfloat[]{\label{fig:RecentZoomed9}
     \includegraphics[width=0.3\linewidth]{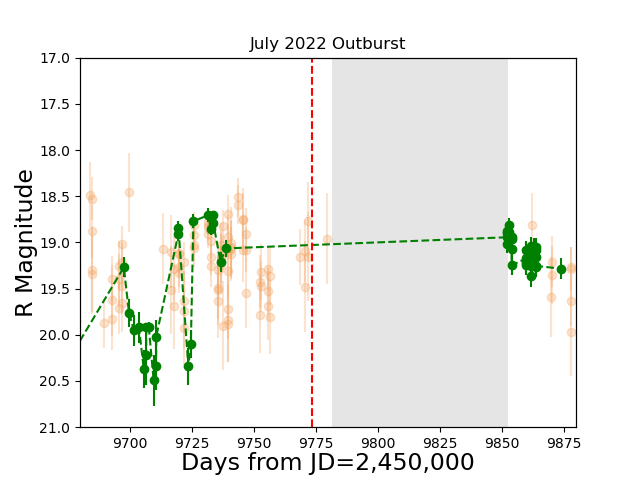}}\\
\end{tabular}
\caption{A zoomed-in light curve of SN~2000ch during its recent outbursts.  We find that SN~2000ch experienced at least nine more outbursts in 2017--2022. Grey dashed curves and circles show the Gaussian comparisons and centre of activity, respectively. The time of the centre of activity is given at the top of each frame. In the case of an outburst with poor data coverage, the predicted time of the centre of activity is displayed instead. Absorption features observed in the light curve might be due to dust formation close to the star. In this scenario, the centre of the activity shows the approximate time of the outburst that might look dimmer owing to dust. The light curve exhibits a wide variety in peak magnitude, duration, and shape of events.}\label{fig:RecentZoomed}
\end{figure*}

{\it January 2017:} Fig.~\ref{fig:RecentZoomed1} shows that in 2017, SN~2000ch faded by $\sim 1$ mag. Owing to a lack of observations before this change, the exact time of the outburst is unclear.  It is likely that the peak was missed.  But, given a period of 200.7 d, the time of the peak brightness is within 61.4 d of an expected event. 

{\it March 2018:} In 2018, SN~2000ch reached a peak of $R= 17.7$ mag. Fig.~\ref{fig:RecentZoomed2} shows SN~2000ch's \nth{16} outburst.  This has better data coverage than the previous one, with a clear rise and decline.

{\it March 2019:} After the sixteenth outburst, in early 2019, SN~2000ch reached a peak at $R= 17.2$ mag, which is as bright as its first outburst in 2000. Fig.~\ref{fig:RecentZoomed3} shows that the \nth{17} outburst has a well-defined rise, and shows multiple oscillations during its decline, similar to some preceding events.

{\it October 2019:} In 2019, SN~2000ch experienced multiple fluctuations in its apparent magnitude, but without a single very well-defined brightness peak (see Fig.~\ref{fig:RecentZoomed4}). These fluctuations resemble secondary peaks observed after previous outbursts, and they start right away when SN~2000ch emerged from behind the Sun. Therefore, it is likely that either the main event took place July--October when SN~2000ch was difficult to observe (shown in grey), or the outburst is fainter than the previous events (or both).

{\it April 2020:} Fig.~\ref{fig:RecentZoomed5} shows that SN~2000ch experienced its \nth{19} outburst in April 2020, where it reached an apparent magnitude of 18.3. This event is similar to the late-2009 outburst, where SN~2000ch reached a similar magnitude of 18.1. Both 2009 and 2020 events are followed by a secondary peak. Later, in Section~\ref{sec:Centroid}, we discuss how we interpret the timing of outbursts that lack a clear single peak, sometimes having two or more comparable peaks. 

{\it November 2020:} SN~2000ch brightened again in late 2020 and reached $R= 17.9$ mag (see Fig.~\ref{fig:RecentZoomed6}). The \nth{20} outburst also has more than one brightness peak, similar to the 2009 and April 2020 events.

{\it June 2021:} In 2021 May, SN~2000ch experienced its \nth{21} documented outburst, shown in Fig.~\ref{fig:RecentZoomed7}. This one is faint compared to many of the other outbursts, with a peak brightness of only $R= 18.7$ mag.  A brighter peak could have been missed, however, if it occurred just prior to this month when there was a lack of data. 

{\it December 2021:} In late 2021, SN~2000ch brightened again ($R \approx 18.1$). The \nth{22} outburst shows a relatively slow rise and decline with a modest peak brightness, but superposed with some rapid flaring (Fig.~\ref{fig:RecentZoomed8}). The main rise and fall is less pronounced than previous events, but the rapid fluctuations are similar. 

{\it July 2022:} In mid 2022, SN~2000ch reached $R= 18.7$ mag (Fig.~\ref{fig:RecentZoomed9}). SN~2000ch brightened twice by more than 1.5 mag. Either the main peak is missing owing to the lack of data or this event is  dominated by the rapid fluctuations similar to the December 2021 outburst.
 
\begin{table*}
\caption{Summary of SN~2000ch's 23 known outbursts from 2000 to 2022.}
\begin{adjustbox}{max width=\textwidth,center}
\renewcommand{\arraystretch}{1.}
\begin{tabular}{ccccccc} 
\hline
Outburst &Peak& Centre of Activity& Approx. Amplitude & Approx. Duration [d]& Ref. &  \\
$\#$ &(JD-2,450,000) & (JD-2,450,000)& &  &  \\
\hline
1&1667.7&1670.2& 3.4& 69.0& \cite{W04}\\
2&3108.8& &2.0 &59.9 &this work \\
3&3741.0& &2.7 & 107.5&this work \\
4&4177.4&&1.2&53.8&this work \\
5&4746.6&4777.9 & 3.6& 64.4& \cite{P10}\\
6&4944.4&4964.6 & 2.3 &69.0 &\cite{P10}\\
7&5163.7&5161.9 & 2.5& 42.8& \cite{P10}\\
8&5392.4& &2.3&138.8& \cite{P13}\\
9&5576.8&&1.2&126.1& \cite{P13}\\
10&5704.4&&1.1&84.9& \cite{P13}\\
11&6013.4&&2.3&148.9& \cite{P13}\\
12&6354.3& & $>$2.4& 114.7 & Nevski et al. (2013)$^a$\\
13&6777.7& &2.0&155.9&this work \\
14&7047.8&&1.6&84.9&this work \\
15&7827.8& & 1.6& 148.0&this work \\
16&8166.9&8184.4 & 2.4& 124.4&this work \\
17&8534.9&8561.8 &3.2& 155.7&this work \\
18&8764.0& & 1.4& 90.9&this work \\
19&8939.7&8959.0 & 2.4& 105.0&this work \\
20&9145.0&9173.8 &2.1 & 101.9&this work \\
21&9341.7& & 2.0& 84.0&this work \\
22&9528.9&&  2.2& 143.8&this work \\
23&9731.7& & 1.8& 41.1&this work \\
\hline
\label{tab:summaryoutbursts}
\end{tabular}
\end{adjustbox}
$^a$V. Nevski, E. Romas, and I. Molotov (CBAT) {\tt \url{http://www.cbat.eps.harvard.edu/unconf/followups/J10524126+3640086.html}}
\end{table*}
\subsection{Centre of Activity vs. Peak Brightness}\label{sec:Centroid}
As we mentioned above, SN~2000ch experienced at least 23 outbursts in 2000--2022. Several more may have occurred at times when SN~2000ch was behind the Sun.  The light curve of SN~2000ch exhibits events with a wide variety in peak magnitude, duration, and sometimes very irregular shape. With multiple repeating events, one is obviously interested to know if these are periodic, and as such, we seek to estimate the timing of events.   However, in some cases it is difficult to give the exact time of each outburst accurately because (1) the time of peak depends on when SN~2000ch is observed and how well the light curve is sampled during that time (i.e., there are significant gaps in the data), and  (2) events can have irregular shapes to the light curves, sometimes with multiple and asymmetric peaks or even eclipse-like events when it becomes dimmer. Given the irregular and multipeaked shapes of various eruptions, we suspect that the time of the brightest recorded magnitude is not necessarily a good measure of the timing of eruptions.   Later we discuss a possible interpretation, where brightening episodes occur around times of periastron in an eccentric massive binary system, and where the irregularity may be caused by erratic instability in one of the two stars (i.e., LBV-like changes in the wind or envelope) or interactions with an asymmetric disk or envelope.  In that case, it may be more useful to characterise a ``centre of activity" for the irregularly shaped light curves of each repeating eruption.  Rather than concentrating on individual brief peaks or valleys, we aim to represent an approximate measure of the overall centre of variation, or the more general trend of brightening and fading.  One may imagine that this is akin to defining the centre of a volcano, which may not be the location with the highest elevation.  This centre of activity is straightforward when a clear rise and decline are seen, but is more uncertain when an event occurs adjacent to a gap in coverage or when there are multiple peaks.

We should also note that the centre of activity is a better measure of the timing of the event if the outbursts are partially blocked out by the star's dust. Absorption dips observed in SN~2000ch's light curve might be caused by extinction from new dust formation, as suggested originally by \cite{W04}. In this paper, we do not investigate dust formation because we only use an R-band filter. More than the R band is required to confirm the change in the color of the star during the dimming event that might be caused by dust formation.

The dashed grey curves in Figs.~\ref{fig:WagnerB}, \ref{fig:PastorelloZoomed} and \ref{fig:RecentZoomed} show Gaussians that represent the overall activity matched to each event that has a well-defined rise and fall. Filled grey circles mark the implied centre of the activity. 
Uncertainties in estimating the centre of the activity are mainly defined based on the lower and upper boundaries of the absorption dips. In Fig.~\ref{fig:WagnerB} and several figures that follow, the red and blue vertical dashed lines show where the centre of the activity is predicted to be using the detected periods of 200.7 and 194 d (see next section for more details), respectively.  The red dashed line is derived from post-2017 events, which are $d_{\rm red} = 1,745.44 + 200.7\,n$. The blue dashed lines are derived from 2000 and 2008--2009 events, which are $d_{\rm blue} = 1,670.16 + 194\,n$. 

In the next section, we search for periodicity in the light curve of SN~2000ch using the time of the outbursts based on peak brightness and also the centre of the activity, plus a Fourier-like estimator and an O--C (observed--calculated) diagram. Finding (quasi)-periodic behaviour could give a hint of the underlying mechanism causing these variations.

\section{Search for Periodicity}\label{sec:Priodicity} 
Previously, the light curve of SN~2000ch had been studied over a decade, but the search for any degree of periodicity was challenging owing to sparse data coverage at that time \citep{P10}. We include our new photometry in the intervening $\sim 12$~yr and we use multiple methods to identify possible periodicity in the light curve of SN~2000ch. In this section, we first use a Fourier-like power-spectrum estimator, and then we further analyse the period variation of the light curve using an O--C diagram.

\begin{figure}
\includegraphics[width=0.45\textwidth]{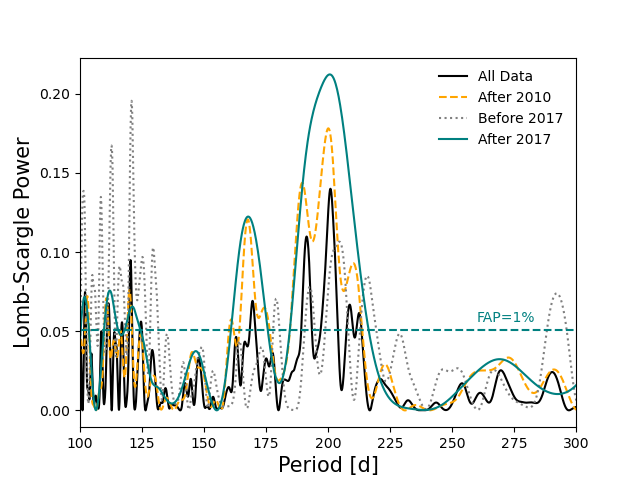}
\caption{The Lomb-Scargle periodogram of SN~2000ch's light curve, where the resulting periodogram depends somewhat on the range of dates used.  Using only the more recent data since 2017 (green), we find a strong but broad peak around 200 d, where the significant width arises because only a few cycles are observed.  Using all the data available from 1994 to 2022 (black solid), or just the data after 2010 (orange dashed), we find narrower peaks, but with the two strongest peaks at 191.4 and 201 d. The double peak comes from early data that are not well sampled and have significant gaps in the time coverage, but may also repeat with a different period.  Nevertheless, a favoured period around 200 d persists regardless of how we divide up the data. The 1\% false alarm probability (FAP) level for the post-2017 periodogram is indicated by a horizontal dashed line.}\label{fig:periodogram}
\end{figure}

We search for periodic variation using a Lomb-Scargle periodogram \citep{L76, S82, v18}, which is designed to detect periodic variations for unequally sampled data. The resulting periodogram is given in Fig.~\ref{fig:periodogram}  for frequencies between 0.003 and 0.01 d$^{-1}$. Using the data, from 1994 to 2022, we find two roughly equal-strength peaks at 191.4 and 201 d.\footnote{We should note that the photometric upper limits are excluded in estimating the periodogram. If we do include the upper limits, the two main peaks will be at 191.3 and 201~d, which are very similar to the results when not including them.}

Detection of multiple peaks in the periodogram may imply that the period of variability is changing over time, and one might wonder if the dip in the middle of the two main peaks might be due to the gaps in the data. To test this idea, we divide the data into two samples: pre-2017 outbursts and post-2017 outbursts. Fig.~\ref{fig:periodogram} shows that the earlier-time baseline gives multiple weaker peaks and no single strong peak because the early data are not well sampled. Using more recent data from 2017 to 2022, however, there is one strong peak at the period of 200.7 d, which is similar to the main peak derived from the full data sample. Therefore, double peaks observed in Fig.~\ref{fig:periodogram} might still be due to the sparse data coverage or might be tracing real differences in the period in the earlier and later times.

The strong peak at the period of 200.7 d has an estimated peak width of $\sim 40$ d. But, unlike a Fourier power spectrum, the width of a peak in the periodogram does not depend on the number of data points and on the signal-to-noise ratio (S/N).  The width of a peak in the periodogram is proportional to the observational baseline, and should not be used to estimate the uncertainty in the period. Therefore, we use the false-alarm probability (FAP) to quantify the significance of a periodogram peak. The highest peak is at 200.7 d, and the FAP of this peak is very small, on the order of $10^{-22}$, which means that this peak is statistically significant.  The horizontal dashed line in Fig.~\ref{fig:periodogram} shows the 1\% FAP level. Based on the duration of the outbursts, the uncertainty in the period should be around 2 d. A shift of 2 d per cycle leads to a 20 d shift over 10 cycles, which is comparable to the width of events. If the time of the outbursts were shifted by more than 2 d per cycle, outbursts would start to shift noticeably. 

\begin{figure*}

\begin{tabular}{ccc}
\subfloat[] {\label{fig:FoldedLC}
    \includegraphics[width=0.3\linewidth]{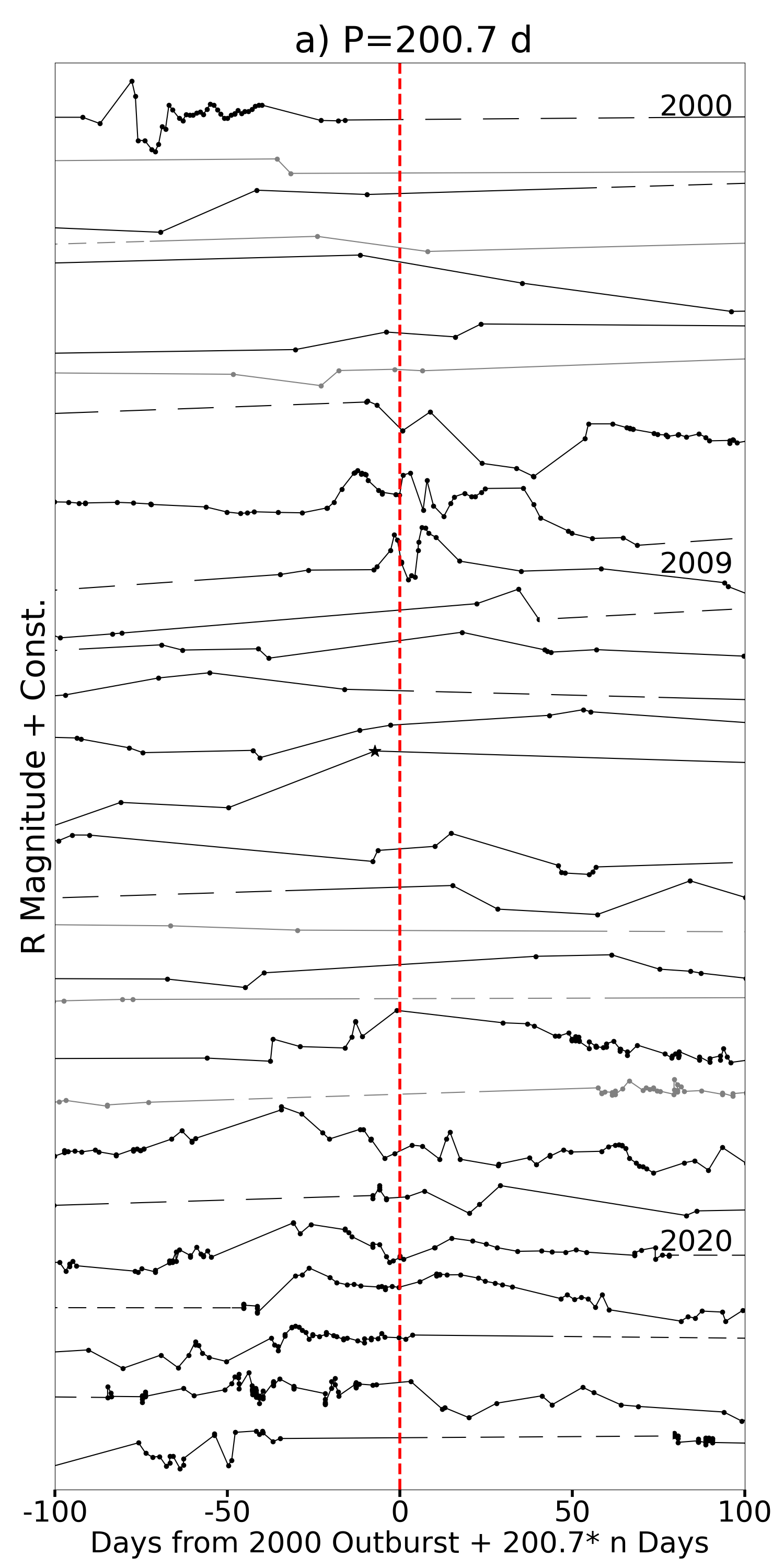}}&
    \subfloat[]{\label{fig:Folded194}
    \includegraphics[width=0.3\linewidth]{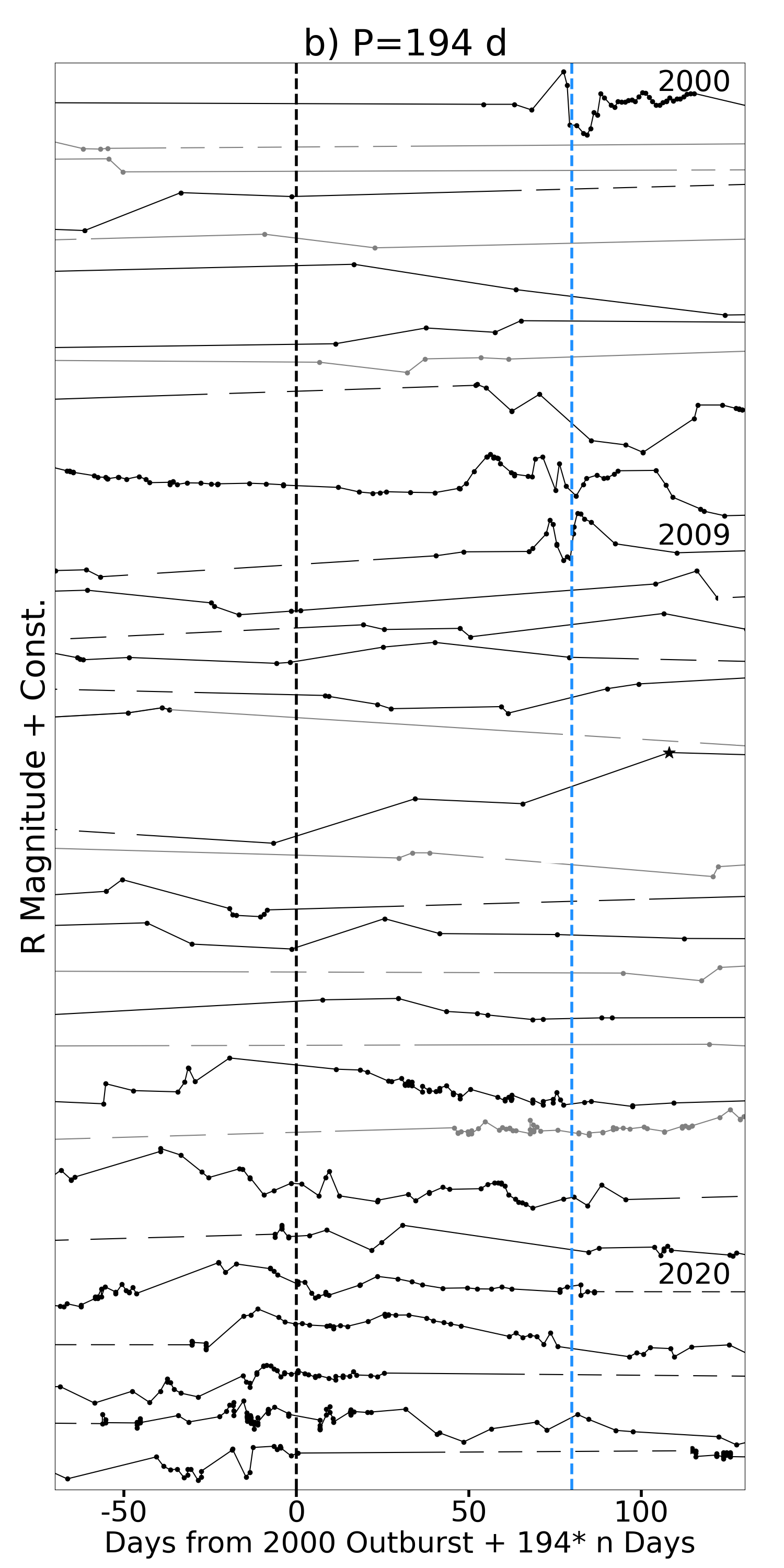}}&
    \subfloat[]{\label{fig:FoldedShorterPeriod}
    \includegraphics[width=0.3\linewidth]{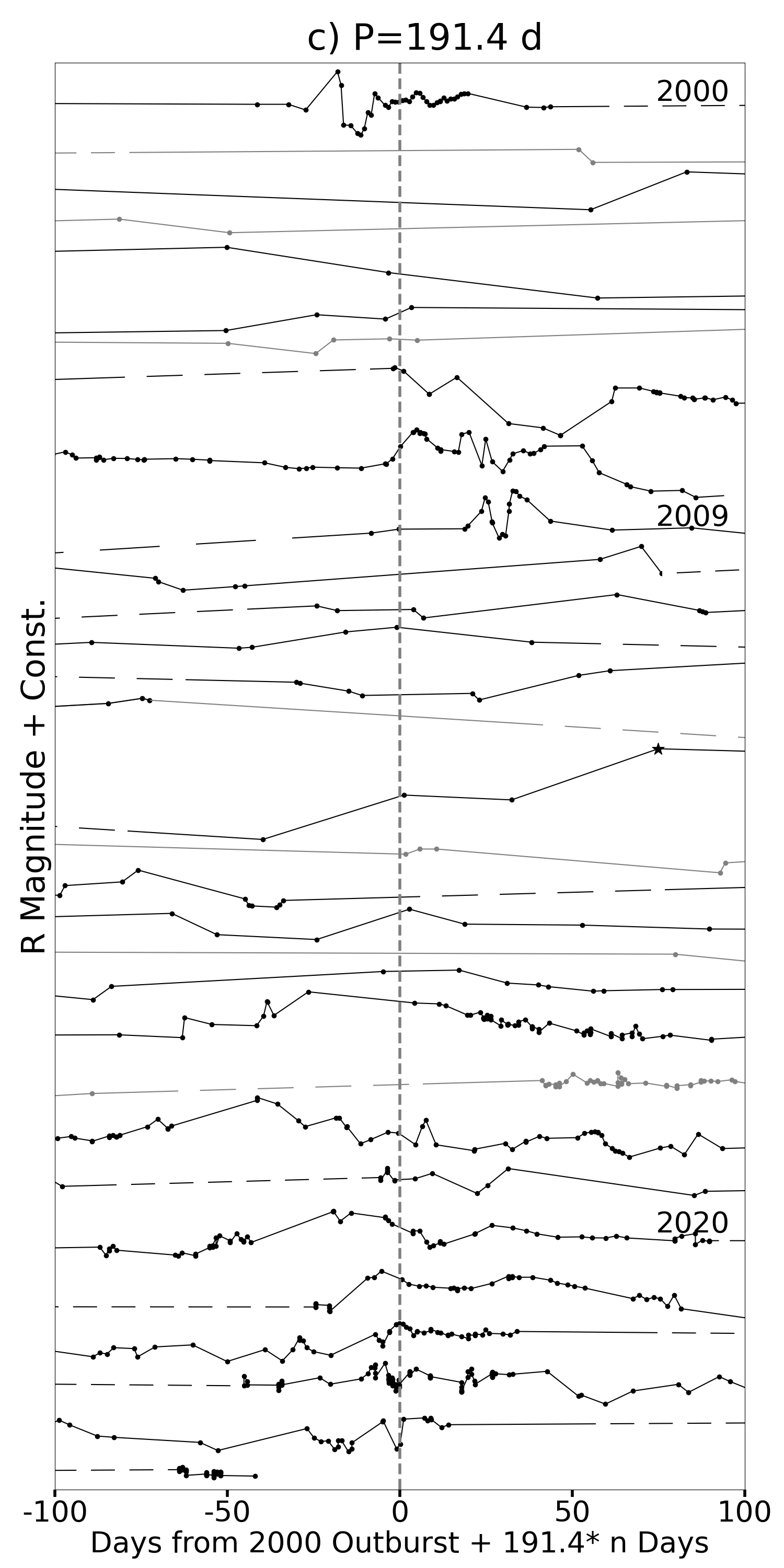}}
\end{tabular}
\caption{Folded light curve of SN~2000ch using three different periods. On the left, the light curve is folded using our preferred period of 200.7 d, estimated from post-2017 outbursts.  ATLAS data and points with upper limits are not shown to reduce the clutter in the figure. Black lines represent the evolution of SN~2000ch during 23 outbursts. Dashed lines denote periods of time when SN~2000ch was difficult to observe during each cycle. The vertical dashed lines show the reference epochs.  Post-2008 Outbursts do repeat with a period of around 200.7 d, but the 2000 outburst favours a shorter period of $\sim 194$ d. The peaks seem to wander when aligning with period of 191.4 d, suggesting that 191.4 d is not a representative period.}\label{fig:FoldedALL}
\end{figure*}

\begin{figure}
\subfloat[]{
  \includegraphics[width=0.45\textwidth]{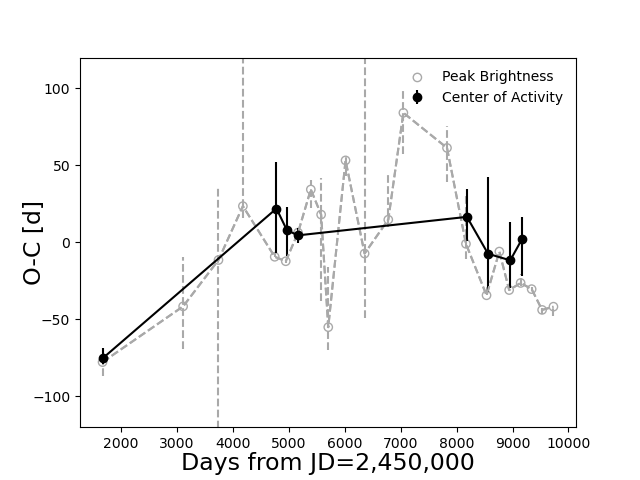}
  \label{fig:OC200}
}
\newline
\subfloat[]{
  \includegraphics[width=0.45\textwidth]{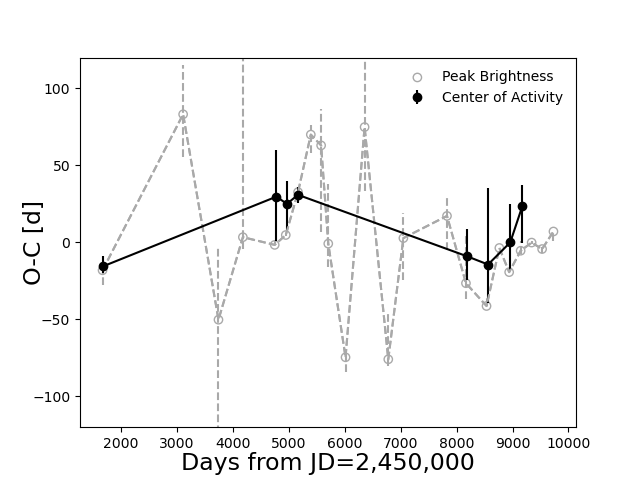}
 \label{fig:OC192}
}
\caption{The top panel (a) shows the observed--calculated (O--C) diagram using a period of 200.7 d. The bottom panel (b) displays the same plot using a period of 191.4 d. The black line shows the O-C diagram using the time of the centre of activity at each outburst (given at the top of the frames in  Figs.~\ref{fig:PastorelloZoomed} and \ref{fig:RecentZoomed}) and the grey dashed line gives the same plot using the time of peak brightness. Using the time of the observed peak brightness, the period of SN~2000ch seems to change during each outburst. However, the observed peak can be shifted from the real peak for a variety of reasons (see text).  On the other hand, when considering the timing of events based on the centre of activity and taking into account the error bars, it appears that the period increased soon after the 2000 event, but has remained relatively unchanged since the 2008/2009 events.}
\label{fig:OCboth}
\end{figure}

\begin{figure*}
\subfloat[]{
  \includegraphics[width=0.58\textwidth]{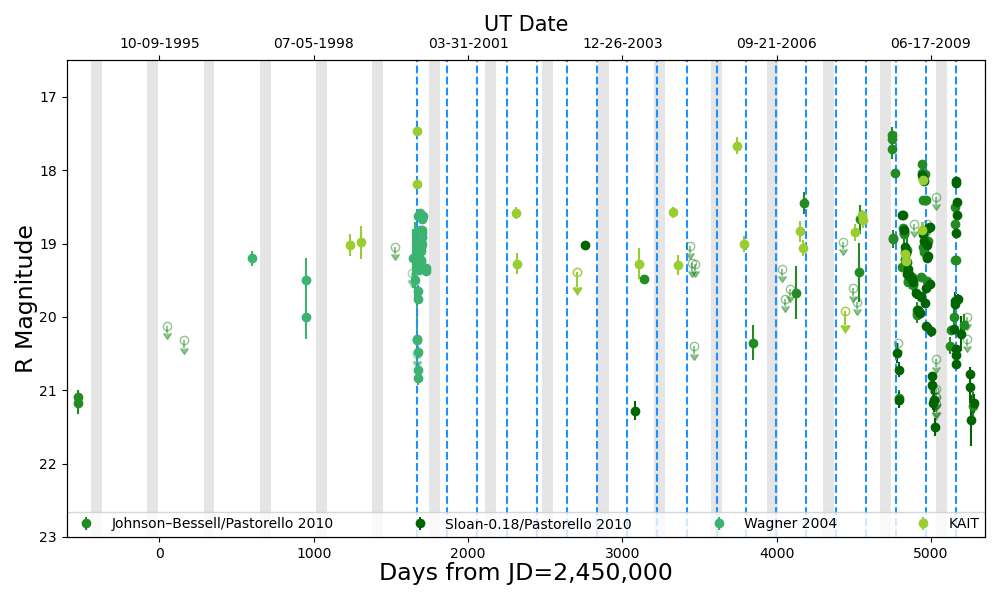}
  \label{fig:LC3A}}\\
\subfloat[]{
  \includegraphics[width=0.58\textwidth]{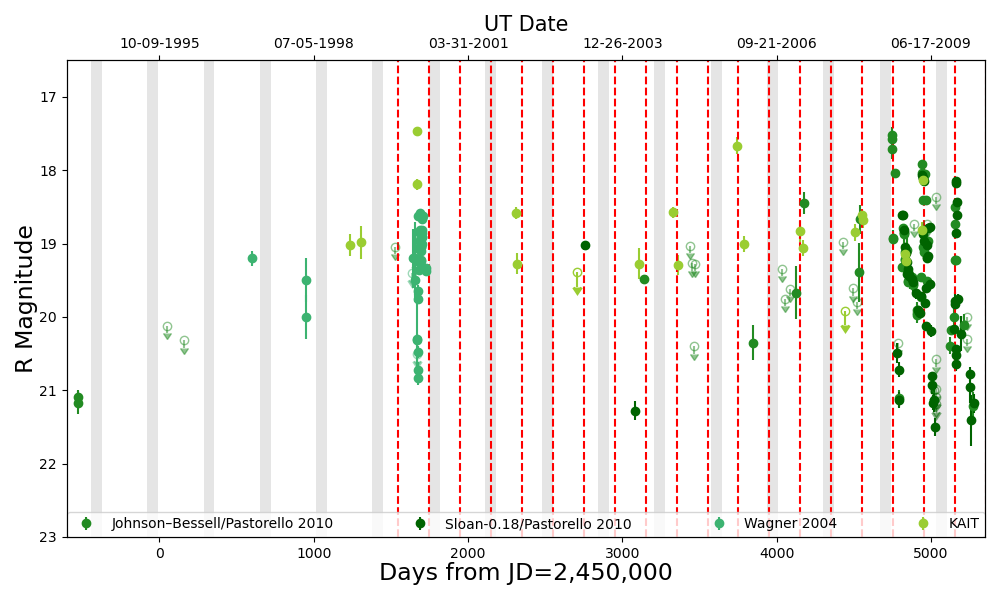}
  \label{fig:LC3B}}\\
  \subfloat[]{
  \includegraphics[width=0.58\textwidth]{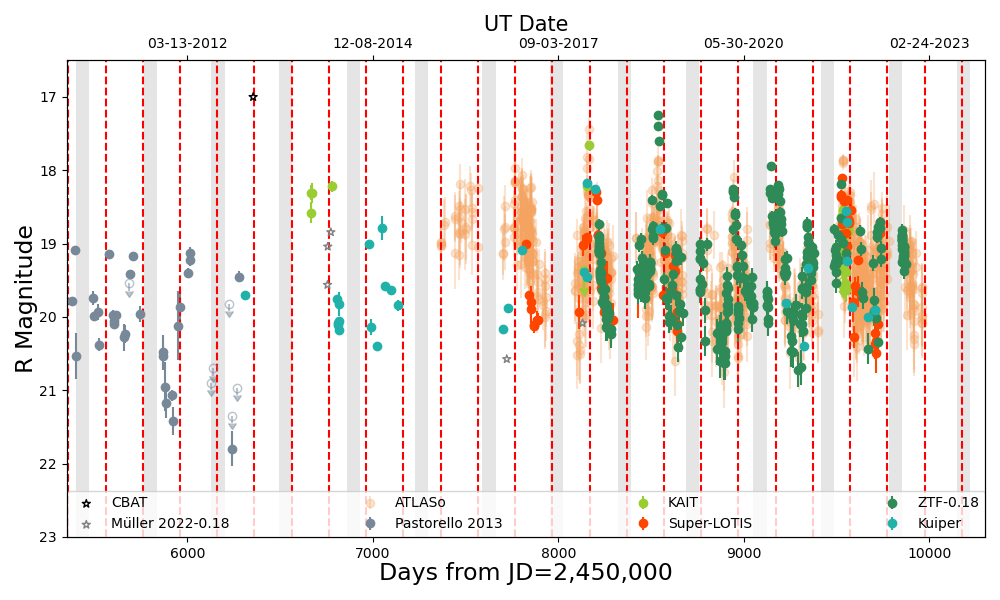}
  \label{fig:LC3C}}
\caption{Same as Fig.~\ref{fig:AllLC}; the top and middle panels show the light curve from 1994 to 2010, and the bottom panel provides the light curve from 2010 to 2022. Red lines in all three panels indicate the reference epoch but using different periods. In the top panel, the reference epoch is the average of the centre of the activities in 2000 and 2008--2009 events using the period of 194 d (similar to the blue dashed line in Fig.~\ref{fig:Folded194}). The middle panel is similar to the top panel but using a longer period of 200.7~d. The shorter period is more consistent with 2000 outburst. The reference epoch in the bottom panel is the average of the centre of activity of post-2017 outbursts using the period of 200.7 d (similar to Fig.~\ref{fig:FoldedLC}). The longer period matches better with the post-2008 outbursts. This suggests that the period is increasing. }\label{fig:LC3}
\end{figure*}

\begin{figure}
 \includegraphics[width=0.45\textwidth]{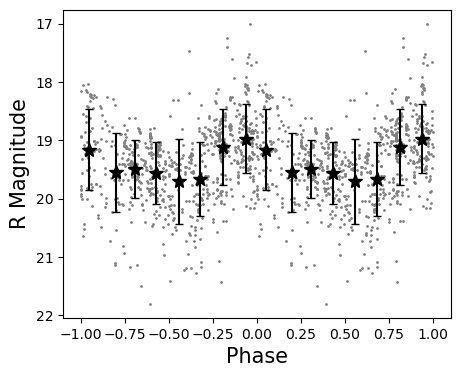}
\caption{Phase diagram using the period of 200.7 d. Black points show average magnitudes after binning the folded light curve.  Using the data from 1994 to 2022, SN~2000ch does seem to get brighter toward phase zero.}\label{fig:phasediagram}
\end{figure}

To further investigate whether the observed peaks correspond to quasiperiodic behaviour of SN~2000ch's outbursts, we fold the light curve with different values for the orbital period.  Fig.~\ref{fig:FoldedALL} shows how the folded light curves look using three different periods of 200.7, 194, and 191.4 days.  Basically, a period of 200.7 days is the best period for the post-2008 data, a period of 194 days seems to provide the best alignment for the 2000 outburst relative to 2008--2009 outbursts, and 191.4 days is a compromise that roughly fits the first event in 2000 and also gives a moderately good alignment of the recent eruptions.  Each of these is discussed below.

{\bf 200.7 days:} Fig.~\ref{fig:FoldedLC} illustrates the light curve folded with a period of 200.7 d, which gives the best alignment of the post-2008 eruptions. All the data are time-shifted $200.7\,n$~d ($n$ is from 0 to 29). Cycles with poor or no data coverage are excluded from Fig.~\ref{fig:FoldedLC}. Dashed portions of the light curve  show months (July--October) when SN~2000ch was was difficult to observe. Black lines represent the evolution of the variable star during the 23 major outbursts. The red vertical dashed line shows the reference epoch. The reference epoch is derived from the average time of the centre of activity in 2017--2022. Quasiperiodic behaviour can be seen from the light curve, because outbursts tend to occur in the middle of the plotted range of time. If the period was regular, one would expect an outburst every $\sim 200.7$ d. Some outbursts are missed either because SN~2000ch's position relative to the Sun, or because there is not enough data coverage at the time of outburst. In other words, when the data are well sampled, there is always an outburst close to the expected time, although there is considerable variation in the shape of the outburst from one cycle to the next.  Some cycles exhibit a clear, well-defined peak, while others have an asymmetric peak, or a combination of multiple peaks and valleys. Thus, as discussed in Section~\ref{sec:Centroid}, we prefer the ``centre of activity" as a better indicator of the timing of events.

{\bf 194 days:} 
As we discussed above, the folded light curve of SN~2000ch seems to show quasiperiodic variability, where the period seen in photometry might be changing over time.
SN~2000ch's first outburst in 2000 better matches the three outbursts in 2008--2009 with a shorter period compared to outbursts after 2017. 
Fig.~\ref{fig:FoldedLC} is effectively like a two-dimensional O--C plot.  In principle, an interacting binary system might evolve toward a longer period if it suffers substantial mass loss during its outbursts.  A shortening period was inferred for this reason befor and after the 19th century eruption of $\eta$ Carinae \citep{sf11}. For example, 5\% mass loss in a short amount of time might increase the period by around 5 d, because of the period's dependence on the mass in Kepler’s Third Law.  If the period is shorter before 2008, then those eruptions should be aligned with a different period.  Fig.~\ref{fig:Folded194} shows the folded light curve with a period approximately between 200.7 and 191.4 d. Using a period of 194 d, three outbursts in 2008--2009 get aligned better with the 2000 event. The black and blue vertical dashed lines indicate the reference epoch using the center of the activity of pre-2010 (2000 and 2008--2009; blue) and post-2017 (2018, 2019, and 2020; black) outbursts, respectively. This better alignment of the earlier eruptions suggests that the period may be increasing over time. Since there are gaps in the data, it is difficult to know if the period change was sudden or gradual.

{\bf 191.4 days:} As we discussed above, using all the data from 1994 to 2022, we find two peaks at 191.4 and 201 d. Fig.~\ref{fig:FoldedShorterPeriod} shows the light curve folded with the shorter period of 191.4 days. Periods of 200.7 and 194 days separately provide good fits to the later and earlier eruptions, respectively, but do a poor job on the other portion of the light curve.  In contrast, a period of 191.4 days seems to be a trade-off in that it roughly fits the first 2000 outburst and the most recent outbursts, but some of the outbursts in between are not fit as well as using either 200.7 or 194 days. If we use this shorter period of 191.4, Fig.~\ref{fig:FoldedShorterPeriod} shows that relative to the 2000 event the period increases and then decreases over time. 
The zero point in this figure is different from the zero point in Fig.~\ref{fig:FoldedLC} 
because it adopts a shorter period.  If a decrease in the period of the variation is real, it could suggest instead that the two stars may be headed toward a merger event owing to loss of angular momentum.

We use the O--C plot as another way to indicate whether SN~2000ch's variations are quasiperiodic. Fig.~\ref{fig:OCboth} shows the O--C diagram using the time of the centre of activity and peak brightness in black and grey dashed lines, respectively. Fig.~\ref{fig:OC200} shows the O--C diagram using a period of 200.7 d, and Fig.~\ref{fig:OC192} displays the same plot using a shorter period of 191.4 d. For a star with no change in the period, one would expect a straight horizontal line across the plot. Using the time of
peak brightness for each outburst, Fig.~\ref{fig:OCboth} would suggest that the period of SN~2000ch seems to be changing over time. However, as noted earlier, this may be a poor way to determine the timing events.  The time of peak can differ significantly from the real peak owing to poor sampling (when peaks are well sampled, they tend to be brief spikes and could be easily missed by gaps in the data).  Moreover, even the real peak can be shifted from the nominal time of the event if dust formation blocks out part of the peak (as in 2000) or if the CSM is inhomogeneous in some scenarios (see Section~\ref{sec:binary}).  As discussed earlier, using the time of the centre of activity to represent the time of the event may be a better way to test for periodicity.  When we examine the times of the centre of activity (black solid lines in Fig.~\ref{fig:OCboth}), we find that adopting the longer period of 200.7 d as the calculated period (panel a) suggests that compared to the 2000 event the observed period is increasing, while adopting the shorter calculated period of 191.4 d (panel b) might imply that the period is decreasing after the 2009 event. However, considering the relatively large error bars, 
the time of the centre of activity after the 2008 outburst, using both periods, does not give a strong indication of a significant change in period.  Continued monitoring of these events with better sampling than in the past may be able to clearly test for the possibility of a changing orbital period that may be an indication of an impending merger event, for example.

Which period --- 200.7, 194, or 191.4 days --- is the correct interpretation of SN~2000ch's repeating eruption? In our assessment, the various representations of the data do seem to favour a period of 194 days for the 2000 outburst, and a longer period of 200.7 days for the eruptions since 2008. Fig.~\ref{fig:LC3} displays the full light curve again with calculated times of events using two different periods of 194 days (blue vertical lines) and 200.7 d (red vertical lines). Figs.~\ref{fig:LC3A} and \ref{fig:LC3B} show the pre-2010 outbursts, where the reference epoch is measured using the period of 194 and 200.7 days respectively. Fig.~\ref{fig:LC3C} shows the post-2010 outbursts, where the reference epoch is measured using the period of 200.7 d. The shorter period clearly matches better with the 2000-2009 outbursts and the longer period matches the outbursts from 2008 to the present. Note, however, that this longer period that matches the later outburst clearly fails at matching the 2000 outburst.   Again, this strongly suggests that the period has increased some time after the 2000 event. For now, SN 2000ch’s light curve is too erratic and multipeaked to constrain the rate of changes in the period precisely, and we cannot determine confidently if the period is slowly and continually changing (as in the prelude to a merger discussed in the previous paragraph) or if there was a one-time shift in the period, as one might expect following a major mass ejection episode in an outburst, for example. Further monitoring and detailed observations may help to constrain the period variation better as additional eruptive events are documented over a longer time baseline and with better sampling.

Finally, Fig.~\ref{fig:phasediagram} illustrates a phase diagram of SN~2000ch, given the period of 200.7 d. The phase diagram shows two complete cycles, where all data points are folded on top of each other twice. To better see the behaviour of SN~2000ch, we bin the folded light curve. Black points in Fig.~\ref{fig:phasediagram} show average magnitudes after binning the folded light curve. SN~2000ch does seem to get brighter toward phase zero.

As previously mentioned, \cite{Muller} also studied the recent variability of SN~2000ch. 
Although their analysis was conducted independently, part of the ZTF dataset used in their analysis was the same as in our work. \cite{Muller} estimated that SN~2000ch outbursts do repeat with a period of 201$\pm$12 d, which is consistent with our estimate of 200.7$\pm$2 d that incorporates a number of additional outbursts.

Above we showed that SN~2000ch experienced multiple outbursts that seem to be periodic or quasiperiodic, perhaps with a change in period, using a variety of ways to test for periodicity.  While the individual eruption events have irregular or even erratic light curves, there does seem to be a clearly repeating periodicity in the approximate centroids of the events. The quasiperiodic variability seen in the light curve of SN~2000ch may be a very important clue to the mechanism that triggers the outbursts. In the next section, we discuss an underlying mechanism that might cause irregular and multipeaked variability with a wide variety of shapes, durations, and magnitudes.

\section{LBV-MODULATED BINARY INTERACTION}\label{sec:binary}
Various clues strongly suggest that the apparent variability of SN~2000ch may arise in an eccentric, interacting binary system.  Specifically, we note the following:

(1) A regular periodicity of 190--200 d implies that there is a binary orbit that governs the repeating events.  Normal S~Dor variability is irregular rather than periodic, and generally exhibits longer timescales.   The period of 190--200 d is short enough that the two massive stars are close to one another and may strongly interact (see below).

(2) The strong periodic variability in brightness suggests that the binary orbit is eccentric.  Whatever specific physical interaction causes the changing brightness (wind collision, accretion, grazing collisions, tidal interaction, mass ejection, etc.), one might naturally expect that the interaction is stronger or more violent when the separation between the two stars periodically decreases at times of periastron encounters.  This is only possible if the orbit is eccentric.

(3) The duration and light-curve shape are somewhat different in each event.  This irregularity in the light curve suggests that at least one of the stars has an instability, causing changes in envelope radius, mass-loss rate, density of circumstellar material, etc., on timescales that are not synchronised with the orbital cycle.  This intrinsic variability may lead to a different intensity and duration of binary interaction at each periastron passage.  Some sort of irregular pulsational or mass-loss variability of one of the stars is perhaps not surprising if the primary star is indeed a massive LBV.

\begin{figure}
  \includegraphics[width=0.45\textwidth]{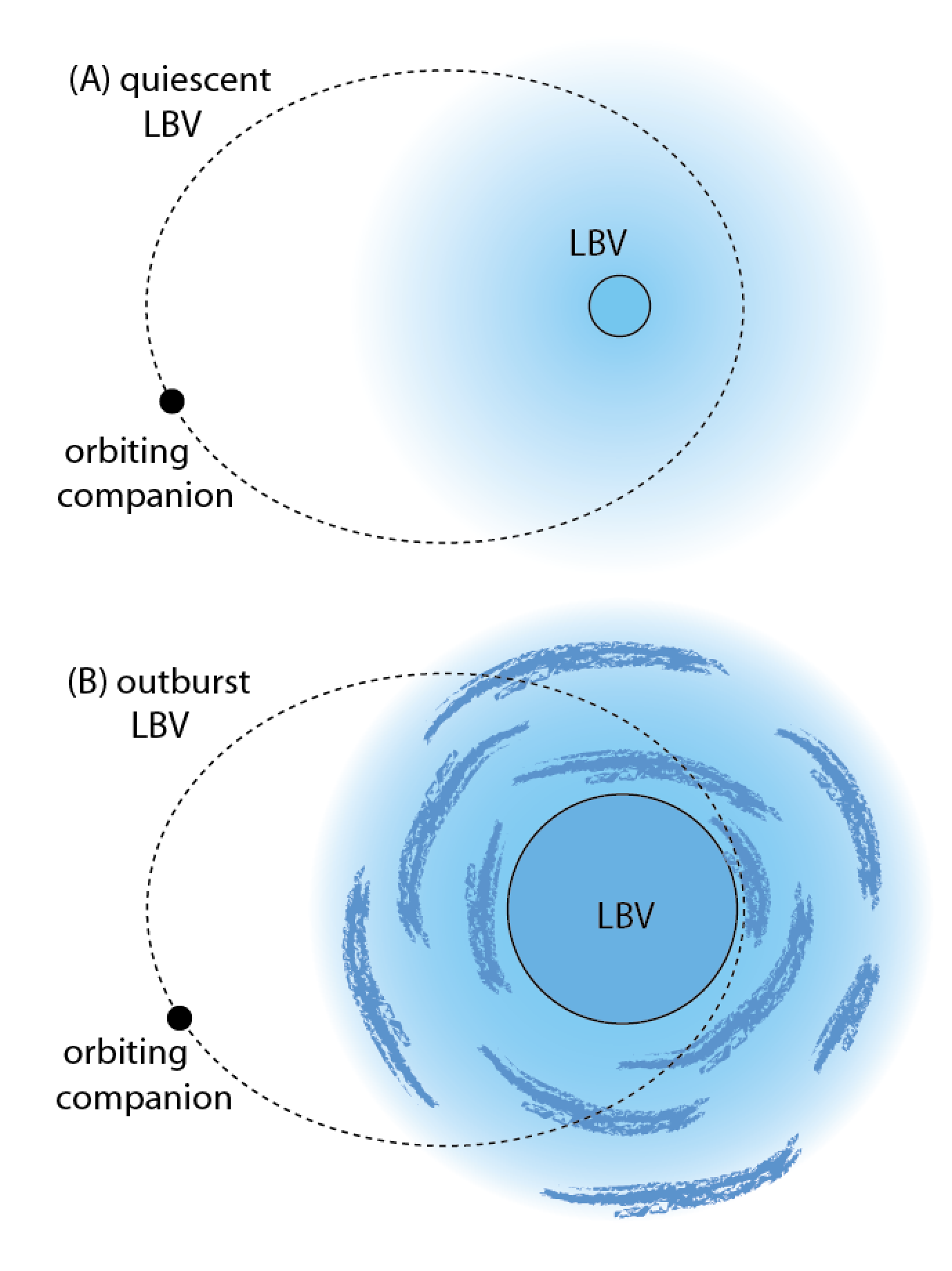}
\caption{Sketch of a possible scenario to explain the periodic $\sim 200$ d outbursts of SN~2000ch.  In an eccentric binary system containing an LBV, the LBV may have intrinsic variability between its normal quiescent and outburst states in its S~Dor cycle. The strength of interaction may vary greatly depending on the state of the LBV at times of closest periastron passage.  The top panel (A) shows a binary system when the LBV is in quiescence, with a comparatively small stellar radius and/or low mass-loss rate.  Any interaction with the companion at periastron may be relatively weak (i.e., normal colliding wind binary) or absent in this case.  The bottom panel (B) shows the same orbital configuration, but with the LBV in its outburst state with an inflated radius and/or higher mass-loss rate that produces dense and possible clumpy CSM.  The blue arcs at various positions are meant to signify clumps or partial shells in the CSM, but the CSM may take the form of spherical shells, spirals, a disk/torus, or a bipolar flow. Much more violent interactions are expected in these periastron passes, with much stronger and highly variable shocks or accretion, or the companion may even temporarily plunge into the inflated envelope of the LBV (depending on the LBV's radius).}\label{fig:orbit}
\end{figure} 

To explore the interaction in such a binary, it is useful to place some rough constraints on the physical parameters of the system.  Even at quiescence, SN~2000ch is very luminous, with an absolute magnitude of $M_R \approx -10.7$ (given the distance of 10.8 $\pm$ 1.2 Mpc adopted by \citealt{P10}). This would be consistent with a very luminous LBV that resides along an evolutionary track for a single star with an initial mass of $\sim 60$~M$_{\odot}$, and a current $L \approx 10^6$~L$_{\odot}$ or perhaps somewhat higher. (Even if some fraction of the quiescent luminosity is due to another persistent source such as an accreting companion, it is still very likely that we are dealing with a high-mass and high-luminosity primary star.)   For a classical high-luminosity LBV \citep{svd04,groh11}, such a star in its hot quiescent state ($T_{\rm eff} \approx 25$~kK) would have a photospheric radius of $\sim 53$~R$_{\odot}$ (0.25 AU), whereas in its cooler outburst state ($T_{\rm eff} \approx 8.5$~kK) at the same bolometric luminosity it may have a much larger radius around 500~R$_{\odot}$ (2.3 AU).  The observed orbital period of $\sim 201$ d then implies a semimajor axis of $a \approx 2.6 \, {\rm AU} \, (M / 60~{\rm M}_{\odot})^{1/3}$.  If the orbit is eccentric ($e \approx 0.5$, for example), then the periaston separation would be about 1.3 AU and the apastron distance about 3.9 AU.  This periastron separation is smaller than the inflated radius of the LBV in its presumed cooler eruptive state (implying very strong interaction), whereas the apastron separation is larger than the stellar radius and thus large enough that we might expect interaction to shut off or weaken at apastron.  This may produce cycles of interaction-driven outburst and quiescence.  The intensity of the periastron interaction would also modulate slowly, depending on the intrinsic state of the LBV.  In some cycles the periastron interaction might be very weak, but at other times we should expect more violent encounters as the LBV attempts to expand.  Perhaps the binary interaction would prevent it from fully reaching its cooler state; that is, it might not get quite as cool as normal LBVs at 8500 K.  It may therefore maintain a warm smooth continuum with strong emission, rather than developing the absorption lines characteristic of an F-type supergiant.

This situation is reminiscent of $\eta$~Car in the lead-up to its 19th century Great Eruption \citep{SN11,S18}, and in the discussion below, we borrow from such a scenario.  $\eta$ Car is known to experience violent wind collisions at periastron passes \citep{pc02,okazaki08}, it has been proposed that accretion onto the companion star may occur and may be related to a variety of observed phenomena \citep{S04, K09}, and it is likely that the companion star plunged into the bloated envelope of the primary before and during the 19th century eruption \citep{SN11,smith18alma,S18}.  SN~2000ch’s outbursts also have similar photometric and spectroscopic properties to the rapid brightening  and fading observed in pre-SN eruptions of SN~2009ip  \citep{S10,P13}. Therefore, the binary interaction of SN~2000ch may be similar to SN~2009ip, as suggested already \citep{P13,S14,S22}.  A binary system at the heart of SN~2000ch might not be too surprising, given recent evidence that most LBVs are the products of binary evolution \citep{ST15,A17}.

Fig.~\ref{fig:orbit} illustrates a cartoon of our proposed scenario to explain SN~2000ch.  This shows the same binary system at two different epochs when the LBV primary is relatively hot in its quiescent state (Fig.~\ref{fig:orbit}a), and at a different epoch when the LBV has expanded during a cool S~Dor-like excursion (Fig.~\ref{fig:orbit}b).  Note that this scenario involves the interplay of two different variability phenomena.  One is the intrinsic ``outburst'' cycle of the LBV, analogous to S~Dor variability and presumably driven by an envelope instability at high luminosity, as in single LBV stars \citep{vg01,svd04,S11,J18}.  The other is the observed ``eruptions'' of SN~2000ch, which are driven by interaction between the two stars at periastron in the eccentric binary system.  The exact mechanism causing the periastron eruptions is uncertain and may be the topic of future study, but it could be caused by wind interaction, grazing collisions, accretion, etc., as noted above.  

The companion star in this scenario may be another massive star or perhaps a compact object. \cite{P10} suggested that the companion star might be a WR star. They observed that during the quiescent state, spectra of SN~2000ch show He~{\sc ii} lines, while they found no evidence for the presence of He~{\sc ii} lines during outburst. However, He~{\sc ii} lines may also naturally disappear if the star's envelope expands and cools in outburst [note that some classic LBVs like AG Car and R127 have a WR-type spectrum in their hot quiescent state \citep{ws82,stahl83,wolf92,groh11}].  He~{\sc ii} emission is also seen from the colliding wind region in $\eta$ Carinae at some phases of its variability \citep{teodoro12}. On the other hand, if the companion is a compact object, then He~{\sc ii} emission may come from an accretion disk, and strongly variable accretion onto a neutron star or black hole may be the most likely power source for SN~2000ch's eruptions.  

A key point is that while periastron eruptions of SN~2000ch should occur with a fairly regular period of $\sim 200$ d, the strength of those interaction-powered eruptions may be strongly modulated by the quiescent or outburst state of the LBV, which is likely to vary on longer timescales of a few to 10~yr.  Because of this, the light curves of periastron eruptions may be quite different from one cycle to the next.  The periastron separation between the centers of the two stars will be the same for each orbital cycle, of course, but the radius of the primary and/or the density and extent of its circumstellar matter (CSM) may vary greatly.  Thus, any component of the luminosity that comes from interaction (which is added to the intrinsic luminosity of the LBV) may also vary greatly.   

We do not know the eccentricity of this orbit yet, but Fig.~\ref{fig:orbit} shows a hypothetical orbit with $e=0.6$ for illustrative purposes; further study may refine this.  With this configuration, the LBV in its hot quiescent state (Fig.~\ref{fig:orbit}a) will have a radius much smaller than the periastron separation, and so interaction may be quite mild (collisions with a relatively low-density wind or a relatively low accretion rate onto the companion, for example).  In this phase, periastron eruptions may be weak or even undetectable in some cycles.   However, when the LBV attempts to expand during an S~Dor cycle, the radius can grow to a size comparable to the periastron separation, at which time the interaction can be catastrophic.  The interaction strength may of course be anywhere in between these two examples in Fig.~\ref{fig:orbit}.  In the orbital configuration shown in Fig.~\ref{fig:orbit}b, the radius of the LBV (attempting to expand to a radius of $\sim 2.6$ AU as noted above) would reach the periastron separation (1.2--1.3 AU) at a temperature of $\sim 11,000$--12,000~K.  Thus, violent interaction like grazing collisions or a brief common-envelope phase would likely prevent the LBV from attaining its fully expanded state, and may trigger significant episodic and asymmetric mass loss.  Over time, this mass loss from the system may cause the orbital period to get longer, as seems to be observed here and as seen in the potentially similar situation of $\eta$~Car \citep{D96,KA10, SN11,sf11}. Such interaction may also limit the temperature decrease, preventing it from reaching the cool $T_{\rm eff} = 8500$~K values expected for a normal S~Dor phase.  We suspect that this is one reason why SN~2000ch may appear quite different from classical LBVs, and why it appears to maintain a hotter spectrum in outburst \citep{S11}.  This may also apply to the unusually hot LBV-like progenitor of SN~2009ip \citep{S10}.  In addition to an expanded stellar radius, the expanded outburst state of the LBV might also have a dense wind, providing denser CSM with which the companion may interact.  An expanded radius or denser CSM at some epochs may translate to brighter observed eruptions of SN~2000ch.

Overall, we propose that the interplay between repeating periastron passes in an eccentric orbit combined with an LBV primary that slowly changes its radius and/or mass-loss properties may provide a plausible explanation for the eruptions of SN~2000ch, which repeat with a fairly regular period of $\sim 200$ d, but exhibit a wide diversity in peak luminosity and light-curve shape from one cycle to the next.  This qualitative scenario for SN~2000ch can be quantitatively constrained and refined with future studies:  analysis of spectra may help constrain the stellar radius, temperature, and mass-loss properties, while observations at other wavelengths such as X-rays may test for colliding wind shocks or evidence of accretion onto a compact companion.  Eccentric orbits will be common among relatively wide massive binaries that experience interaction late in their evolution, so one may expect SN~2000ch-like systems to contribute significantly to the observed transient population \citep{SN11}.  This basic scenario can also be adapted to explain other eruptive systems (such as the LBV progenitor of SN~2009ip or others) by altering the period, eccentricity, and masses of the stars (and possibly also the nature of the companion).

\section{Conclusion}\label{sec:consclusion}
We present new photometric observations of SN~2000ch using multiple facilities. This target was already known to have experienced four outbursts in 2000--2010 \citep{W04, P10}. Observations now reveal that SN~2000ch experienced three more outbursts in 2004--2007, and sixteen more outbursts in 2010--2022 with a wide variety of shapes, durations, and magnitudes. The goal of this paper is to explore the possible periodicity of these repeating eruptive events to better understand the underlying mechanism that triggers them.

Our investigation shows that the eruptive outbursts do repeat with a period of around $200.7\pm{2}$ d, but that the first detected eruption in 2000 may favour a shorter period of $\sim 194$ d around that time. Quasiperiodicities observed in the light curve along with diversity in the properties of outbursts lead us to propose a scenario where the underlying mechanism for the outbursts is violent encounters at periastron in a binary system containing an LBV.  If the LBV primary star experiences S~Dor-like episodes, the change in radius and mass-loss rate may lead to irregular eruptions similar to events observed in SN~2000ch's light curve, which are somewhat different in each periastron event. In this scenario, if the primary star inflates significantly during an orbit and therefore has an extended envelope before the next periastron encounter, the companion star may crash through the envelope in a brief common-envelope event. Binary encounters at periastron may also exhibit quasiperiodic behaviour or may even appear to skip some cycles because, at some times of periastron, the primary star may be in a quiescent state with a relatively compact radius.  
While mass transfer in circularised close massive binaries is already a sufficiently difficult theoretical problem, the added complications of eccentric orbits and variable primary stars is likely to occur in nature and may contribute significantly to the observed population of transient sources.  Theoretical work on such systems is warranted.

SN~2000ch will likely experience more outbursts in the future, which may look different from previous events as a result of interactions with its unstable companion during periastron. Given the detected period of 200.7 d in recent eruptive events, we predict that SN~2000ch's next outburst will be around Jan./Feb. 2023 and then again in August 2023 and March 2024, although the event in August 2023 is likely to be unobservable. It is important to mention that the outburst in Jan./Feb. 2023 coincides in time with the resubmission of this work. However, the data covering this outburst are not yet publicly accessible, so we leave the current event to a future study. It would be worthwhile to monitor SN~2000ch before the proposed times of outburst because SN~2000ch's light curve shows that the outbursts sometimes begin months before the nominal centre of activity, and the period may be changing over time. 
Long-term monitoring of SN~2000ch is also of considerable interest in case these events are a prelude to a stellar merger or an SN explosion.

\section*{Acknowledgements}
This research has made use of the NASA/IPAC Infrared Science Archive, which is funded by the National Aeronautics and Space Administration (NASA) and operated by the California Institute of Technology. This work has made use of data from the Asteroid Terrestrial-impact Last Alert System (ATLAS) project. The ATLAS project is primarily funded to search for near-Earth objects (NEOs) through NASA grants NN12AR55G, 80NSSC18K0284, and 80NSSC18K1575; byproducts of the NEO search include images and catalogues from the survey area. This work was partially funded by Kepler/K2 grant J1944/80NSSC19K0112, STFC grants ST/T000198/1 and ST/S006109/1, and HST grants AR-14316 and GO-15889 from the Space Telescope Science Institute (STScI), which is operated by AURA, Inc., under NASA contract NAS5-26555. The ATLAS science products have been made possible through the contributions of the University of Hawaii Institute for Astronomy, the Queen’s University Belfast, the Space Telescope Science Institute, the South African Astronomical Observatory, and The Millennium Institute of Astrophysics (MAS), Chile. We thank the following U.C. Berkeley undergraduate students, as well as Kelsey I.\ Clubb, for assistance with the Lick/Nickel observations: Kyle Blanchard, James Bradley, Chadwick Casper, Daniel P. Cohen, Kiera Fuller, Jenifer Gross, Minkyu Kim, Philip Lu, and Heechan Yuk. A.V.F.'s SN group at U.C. Berkeley has received generous financial assistance from the Christopher R. Redlich Fund and numerous individual donors. KAIT and its ongoing operation were made possible by donations from Sun Microsystems, Inc., the Hewlett-Packard Company, AutoScope Corporation, Lick Observatory, the U.S. NSF, the University of California, the Sylvia \& Jim Katzman Foundation, and the TABASGO Foundation.  Research at Lick Observatory is partially supported by a generous gift from Google, Inc. J.E.A.\ is supported by the international Gemini Observatory, a program of NSF's NOIRLab, which is managed by the Association of Universities for Research in Astronomy (AURA) under a cooperative agreement with the National Science Foundation, on behalf of the Gemini partnership of Argentina, Brazil, Canada, Chile, the Republic of Korea, and the United States of America.

\section*{Data Availability}
The Super-LOTIS, KAIT, and Kuiper data used in this work are available in the article. The ZTF data are available in the public domain: \url{https://irsa.ipac.caltech.edu/Missions/ztf.html}. The ATLAS data are available in the ATLAS Forced Photometry server: \url{https://fallingstar-data.com/forcedphot/}.

\bibliographystyle{mnras}
\DeclareRobustCommand{\VAN}[3]{#3}
\bibliography{MA}

\appendix
\section{The photometric data}
The Super-LOTIS, KAIT, and Kuiper data used in this work are presented in Table~\ref{tab:KAIT}, \ref{tab:SuperLOTIS}, and \ref{tab:Kuiper}.

\label{sec:appendix}
\begin{table}
\caption{Unfiltered KAIT Photometry (roughly $R$ band).}\tiny
\centering
\label{tab:KAIT}
\begin{tabular}{lccccc} 
\hline
MJD& Mag& $-\sigma$ & $+\sigma$ & Lim~mag \\
\hline
51233.3611&19.02&18.87&19.17&19.29\\
51306.1708&18.98&18.75&19.20&19.15\\
51667.2281&17.46&17.42&17.51&19.49\\
51668.2670&18.19&18.11&18.26&19.35\\
52311.3896&18.58&18.50&18.67&19.60\\
52315.4293&19.27&19.13&19.42&19.46\\
52707.3181&19.39&19.19&19.59&19.26\\
53108.2566&19.27&19.07&19.48&19.29\\
53326.5547&18.57&18.50&18.65&19.37\\
53358.5206&19.30&19.16&19.43&19.66\\
53740.4845&17.66&17.55&17.78&18.87\\
53787.3975&19.00&18.90&19.11&19.75\\
54149.3871&18.83&18.69&18.98&19.11\\
54169.3417&19.07&18.95&19.18&19.50\\
54444.5665&19.92&19.73&20.12&19.81\\
54506.4242&18.85&18.73&18.96&19.24\\
54553.2961&18.61&18.55&18.67&19.74\\
54561.3150&18.68&18.57&18.79&19.44\\
54832.5282&19.14&19.02&19.25&19.69\\
54838.4547&19.23&19.07&19.40&19.40\\
54939.2707&18.81&18.71&18.91&19.39\\
54946.3075&18.14&18.06&18.21&19.02\\
56663.4416&18.58&18.44&18.72&19.13\\
56667.4682&18.30&18.20&18.41&19.02\\
56672.3877&18.31&18.17&18.44&18.73\\
56777.2448&18.22&18.15&18.29&18.90\\
58139.5253&19.36&19.12&19.61&19.21\\
58154.5145&18.21&18.14&18.29&19.15\\
58166.4014&17.66&17.60&17.72&19.36\\
59525.5686&18.63&18.52&18.75&18.89\\
59525.5695&18.88&18.67&19.10&19.00\\
59529.5264&18.87&18.70&19.04&18.60\\
59529.5272&19.08&18.89&19.26&18.47\\
59529.5280&18.90&18.74&19.05&18.63\\
59529.5396&18.91&18.81&19.00&19.43\\
59529.5404&19.01&18.91&19.10&19.50\\
59529.5413&18.95&18.88&19.03&19.54\\
59529.5421&18.88&18.80&18.96&19.41\\
59529.5429&18.93&18.83&19.03&19.52\\
59529.5441&18.92&18.84&19.00&19.51\\
59529.5450&18.99&18.90&19.08&19.49\\
59529.5458&18.94&18.86&19.02&19.45\\
59529.5466&19.00&18.90&19.09&19.49\\
59529.5475&18.87&18.79&18.96&19.53\\
59529.5486&18.98&18.89&19.08&19.56\\
59529.5494&18.95&18.86&19.04&19.47\\
59529.5503&19.16&19.07&19.26&19.54\\
59529.5511&19.07&18.98&19.17&19.45\\
59529.5519&18.93&18.83&19.02&19.51\\
59530.5337&18.97&18.88&19.06&19.29\\
59530.5345&19.21&19.11&19.31&19.60\\
59530.5354&19.23&19.15&19.32&20.02\\
59530.5362&19.30&19.21&19.39&19.93\\
59530.5370&19.09&19.01&19.18&19.96\\
59531.5492&19.24&19.13&19.36&19.56\\
59531.5500&19.58&19.44&19.73&19.99\\
59531.5508&19.35&19.24&19.46&19.59\\
59531.5517&19.24&19.15&19.33&19.91\\
59531.5525&19.32&19.21&19.43&19.63\\
59532.5176&19.50&19.38&19.63&19.43\\
59532.5184&19.38&19.26&19.49&19.79\\
59532.5192&19.26&19.14&19.37&19.73\\
59535.5478&18.60&18.52&18.67&19.70\\
59535.5486&18.56&18.50&18.61&19.91\\
59535.5494&18.64&18.56&18.72&19.86\\
59535.5503&18.54&18.48&18.60&19.50\\
59535.5511&18.75&18.67&18.83&19.88\\
59541.5118&18.78&18.68&18.88&19.23\\
59541.5126&18.78&18.67&18.89&19.17\\
59541.5134&18.89&18.79&19.00&19.12\\
59541.5143&18.89&18.78&19.00&19.42\\
59541.5151&18.82&18.71&18.92&19.41\\
59550.5593&19.62&19.53&19.72&20.06\\
59550.5601&19.55&19.43&19.66&20.03\\
59550.5609&19.39&19.28&19.50&19.99\\
59550.5618&19.63&19.46&19.79&19.99\\
59550.5626&19.60&19.47&19.72&19.92\\
59559.5117&18.58&18.52&18.65&19.48\\
59559.5125&18.76&18.69&18.83&19.51\\
59559.5133&18.82&18.74&18.90&19.52\\
59559.5141&18.77&18.69&18.85&19.31\\
59559.5150&18.85&18.77&18.92&19.61\\

\hline
\end{tabular}
\end{table}

\begin{table}
\caption{Super-LOTIS Photometry Data.}\scriptsize
\centering
\label{tab:SuperLOTIS}
\begin{tabular}{lccccccc} 
\hline
JD$-$2,456,000& R & $\sigma_{\rm{R}}$ & V &$\sigma_{\rm{V}}$  & I & $\sigma_{\rm{I}}$ \\
\hline
1705.00&&&20.63&0.23&&&\\
1763.90&&&18.65&0.04&&&\\
1806.80&&&19.10&0.18&&&\\
1827.80&19.01&0.06&&&&\\
1841.70&19.70&0.06&&&&\\
1850.70&19.80&0.22&&&&\\
1853.70&19.89&0.19&&&&\\
1866.70&20.13&0.09&&&&\\
1869.70&20.10&0.09&&&&\\
1886.70&20.04&0.13&&&&\\
1889.70&20.04&0.09&&&&\\
2112.00&19.94&0.23&20.55&0.20&&\\
2118.80&&&&&20.03&0.27\\
2119.90&&&21.20&0.49&&\\
2124.90&&&&&18.32&0.27\\
2131.00&19.02&0.14&19.58&0.11&&\\
2153.90&18.94&0.05&&&&\\
2156.90&18.91&0.04&&&&\\
2204.80&18.30&0.04&&&&\\
2206.80&18.40&0.06&&&&\\
2212.90&18.88&0.05&&&&\\
2213.90&18.89&0.07&&&&\\
2217.70&19.03&0.06&&&&\\
2222.70&19.46&0.07&&&&\\
2224.80&19.39&0.12&&&&\\
2229.80&19.13&0.08&&&&\\
2231.70&19.60&0.13&&&&\\
2236.70&19.32&0.21&&&&\\
2244.60&19.73&0.06&&&&\\
2246.70&19.93&0.17&&&&\\
2248.60&19.90&0.21&&&&\\
2261.70&19.48&0.24&&&&\\
2295.60&20.04&0.23&&&&\\
2431.00&19.67&0.34&20.48&0.32&19.21&0.36\\
2560.80&18.85&0.08&&&&\\
2564.80&19.70&0.14&&&&\\
2575.80&19.13&0.04&&&&\\
2612.70&19.62&0.07&&\\
2616.70&19.30&0.15&&\\
2627.70&19.38&0.08&&\\
2629.70&19.15&0.06&&\\
2631.70&19.08&0.07&&\\
2637.70&19.93&0.26&&\\
2638.70&20.09&0.21&&\\
2640.70&20.20&0.21&&\\
3524.90&18.33&0.06&&\\
3525.90&18.36&0.04&&\\
3528.90&18.11&0.03&&\\
3529.90&18.73&0.10&19.36&0.09\\
3530.90&18.87&0.05&&\\
3537.90&18.42&0.11&&\\
3552.90&18.86&0.12&&\\
3553.90&18.40&0.03&&\\
3554.90&19.04&0.11&19.65&0.10\\
3559.90&18.67&0.04&&\\
3560.90&18.67&0.08&19.22&0.07\\
3564.90&18.72&0.14&19.21&0.11\\
3565.90&18.70&0.05&&\\
3575.90&18.54&0.11&&\\
3585.80&19.80&0.27&20.35&0.15\\
3592.80&20.27&0.14&&\\
3600.80&19.58&0.12&&\\
3613.90&19.23&0.06&&\\
3616.90&19.65&0.21&19.65&0.17\\
3705.70&20.37&0.20&&\\
3706.70&20.22&0.32&&\\
3709.70&20.49&0.27&&\\
3710.70&20.33&0.26&&\\
3724.70&20.10&0.15&&\\
3732.70&18.86&0.06&&\\
3733.60&18.71&0.04&&\\

\hline
\end{tabular}
\end{table}

\begin{table}
\caption{Kuiper/MONT4K Photometry Data.}\scriptsize
\centering
\label{tab:Kuiper}
\begin{tabular}{lccc} 
\hline
JD& Harris-R & $\sigma_{\rm{R}}$ \\
\hline

2456311.80&19.71&0.02\\
2456808.80&19.75&0.03\\
2456809.80&20.09&0.03\\
2456810.80&20.12&0.04\\
2456817.80&20.18&0.06\\
2456818.80&20.05&0.03\\
2456819.80&19.83&0.16\\
2456978.90&19.01&0.03\\
2456992.00&20.14&0.11\\
2457020.90&20.40&0.04\\
2457047.80&18.79&0.16\\
2457063.80&19.58&0.03\\
2457097.90&19.63&0.04\\
2457134.70&19.84&0.08\\
2457699.00&20.17&0.04\\
2457727.01&19.87&0.03\\
2457805.84&19.08&0.02\\
2458138.94&19.39&0.02\\
2458151.89&19.45&0.04\\
2458154.89&18.17&0.01\\
2458197.71&18.25&0.01\\
2458548.77&18.80&0.01\\
2459227.99&19.81&0.02\\
2459321.74&20.39&0.03\\
2459345.68&19.33&0.01\\
2459552.93&18.55&0.01\\
2459553.96&18.70&0.01\\
2459554.97&19.23&0.01\\
2459585.01&19.86&0.02\\
2459666.68&20.00&0.04\\
2459706.73&19.91&0.04\\
2459707.72&19.91&0.05\\

\hline
\end{tabular}
\end{table}
\label{lastpage}
\end{document}